\documentclass[apjl]{emulateapj}
\usepackage{comment}
\usepackage{ifthen}

\newcommand{\forloop}[5][1]%
{%
\setcounter{#2}{#3}%
\ifthenelse{#4}%
	{%
	#5%
	\addtocounter{#2}{#1}%
	\forloop[#1]{#2}{\value{#2}}{#4}{#5}%
	}%
	{%
	}%
}%

\newcommand{\ctbd}[1]{}

\newcommand{\lc}{light curve}
\newcommand{\lcs}{light curves}
\newcommand{\Lc}{Light curve}

\newcommand{\band}[1]{\ensuremath{#1}~band}

\newcommand{\kms}{\ensuremath{\rm km\,s^{-1}}}
\newcommand{\ms}{\ensuremath{\rm m\,s^{-1}}}

\newcommand{\gcmc}{\ensuremath{\rm g\,cm^{-3}}}
\newcommand{\ergscmsq}{\ensuremath{\rm erg\,s^{-1}\,cm^{-2}}}

\newcommand{\vsini}{\ensuremath{v \sin{i}}}
\newcommand{\feh}{\ensuremath{\rm [Fe/H]}}

\newcommand{\vmac}{\ensuremath{v_{\rm mac}}}
\newcommand{\vmic}{\ensuremath{v_{\rm mic}}}

\newcommand{\rsun}{\ensuremath{R_\sun}}
\newcommand{\msun}{\ensuremath{M_\sun}}
\newcommand{\lsun}{\ensuremath{L_\sun}}

\newcommand{\rstar}{\ensuremath{R_\star}}
\newcommand{\mstar}{\ensuremath{M_\star}}
\newcommand{\lstar}{\ensuremath{L_\star}}

\newcommand{\teffstar}{\ensuremath{T_{\rm eff\star}}}
\newcommand{\rhostar}{\ensuremath{\rho_\star}}
\newcommand{\loggstar}{\ensuremath{\log{g_{\star}}}}

\newcommand{\rpl}{\ensuremath{R_{p}}}
\newcommand{\mpl}{\ensuremath{M_{p}}}

\newcommand{\rhopl}{\ensuremath{\rho_{p}}}

\newcommand{\arstar}{\ensuremath{a/\rstar}}
\newcommand{\zrstar}{\ensuremath{\zeta/\rstar}}

\newcommand{\rjup}{\ensuremath{R_{\rm J}}}
\newcommand{\mjup}{\ensuremath{M_{\rm J}}}

\newcommand{\refsec}[1]{\mbox{\S\ \ref{sec:#1}}}

\newcommand{\reffigl}[1]{Figure~\ref{fig:#1}}
\newcommand{\refsecl}[1]{\mbox{Section \ref{sec:#1}}}

\newcommand{\reftabl}[1]{Table~\ref{tab:#1}}

\newcommand{\flwof}{\mbox{FLWO 1.2\,m}}

\newcommand{\flwos}{\mbox{FLWO 1.5\,m}}

\newcommand{\hatcurhtr}{HTR315-006}                                    
\newcommand{\hatcurfield}{314}                                         
\newcommand{\hatcurCCra}{\ensuremath{07^{\mathrm h}15^{\mathrm m}18.00{\mathrm s}}}                                  
\newcommand{\hatcurCCdec}{\ensuremath{+14{\arcdeg}15{\arcmin}45.4{\arcsec}}}                                 
\newcommand{\hatcurCCtwomass}{2MASS~07151801+1415453}                  
\newcommand{\hatcurCCgsc}{GSC~0774-01441}                              
\newcommand{\hatcurCCtassmv}{11.818}                                   
\newcommand{\hatcurCCtwomassJmag}{\ensuremath{10.797\pm0.022}}         
\newcommand{\hatcurCCtwomassHmag}{\ensuremath{10.589\pm0.024}}         
\newcommand{\hatcurCCtwomassKmag}{\ensuremath{10.543\pm0.020}}         
\newcommand{\hatcurCCesoJKmag}{\ensuremath{0.274\pm0.032}}             
\newcommand{\hatcurLCdip}{\ensuremath{7.2}}                            
\newcommand{\hatcurLCrprstar}{\ensuremath{0.0970\pm0.0012}}            
\newcommand{\hatcurLCbsq}{\ensuremath{0.036_{-0.021}^{+0.042}}}        
\newcommand{\hatcurLCimp}{\ensuremath{0.189_{-0.080}^{+0.083}}}        
\newcommand{\hatcurLCzeta}{\ensuremath{14.31\pm0.06}}                  
\newcommand{\hatcurLCdur}{\ensuremath{0.1539\pm0.0008}}                
\newcommand{\hatcurLCingdur}{\ensuremath{0.0141\pm0.0006}}             
\newcommand{\hatcurLCP}{\ensuremath{3.355240\pm0.000007}}              
\newcommand{\hatcurLCPprec}{\ensuremath{3.3552401}}                    
\newcommand{\hatcurLCPshort}{\ensuremath{3.3552}}                      
\newcommand{\hatcurLCT}{\ensuremath{2455216.97667\pm0.00028}}          
\newcommand{\hatcurLCTA}{\ensuremath{2454405.00857\pm0.00162}}         
\newcommand{\hatcurLCTB}{\ensuremath{2455243.81859\pm0.00029}}         
\newcommand{\hatcurLCiblendA}{\ensuremath{0.83\pm0.05}}                
\newcommand{\hatcurLCiblendB}{\ensuremath{0.69\pm0.03}}                
\newcommand{\hatcurSMEiteff}{\ensuremath{6188\pm80}}                   
\newcommand{\hatcurSMEizfeh}{\ensuremath{-0.26\pm0.08}}                
\newcommand{\hatcurSMEizfehshort}{\ensuremath{-0.26}}                  
\newcommand{\hatcurSMEilogg}{\ensuremath{4.01\pm0.06}}                 
\newcommand{\hatcurSMEivsin}{\ensuremath{10.5\pm0.5}}                  
\newcommand{\hatcurSMEivmac}{\ensuremath{4.66}}                        
\newcommand{\hatcurSMEivmic}{\ensuremath{0.85}}                        
\newcommand{\hatcurSMEiiteff}{\ensuremath{6373\pm80}}                  
\newcommand{\hatcurSMEiizfeh}{\ensuremath{-0.16\pm0.08}}               
\newcommand{\hatcurSMEiizfehshort}{\ensuremath{-0.16}}                 
\newcommand{\hatcurSMEiilogg}{\ensuremath{4.25\pm0.06}}                
\newcommand{\hatcurSMEiivsin}{\ensuremath{10.0\pm0.5}}                 
\newcommand{\hatcurSMEiivmac}{\ensuremath{4.94}}                       
\newcommand{\hatcurSMEiivmic}{\ensuremath{0.85}}                       
\newcommand{\hatcurDSteff}{\ensuremath{7000\pm100}}                    
\newcommand{\hatcurDSlogg}{\ensuremath{4.5\pm0.25}}                    
\newcommand{\hatcurDSvsini}{\ensuremath{11.2\pm1.0}}                   
\newcommand{\hatcurDSgamma}{\ensuremath{-2.09\pm0.74}}                 
\newcommand{\hatcurDSrvrms}{\ensuremath{0.74}}                         
\newcommand{\hatcurLBii}{\ensuremath{0.1858}}                          
\newcommand{\hatcurLBiii}{\ensuremath{0.3625}}                         
\newcommand{\hatcurISOmlong}{\ensuremath{1.191\pm0.042}}               
\newcommand{\hatcurISOrlong}{\ensuremath{1.317\pm0.068}}               
\newcommand{\hatcurISOlogg}{\ensuremath{4.27\pm0.04}}                  
\newcommand{\hatcurISOlum}{\ensuremath{2.56\pm0.31}}                   
\newcommand{\hatcurISOmv}{\ensuremath{3.74\pm0.14}}                    
\newcommand{\hatcurISOage}{\ensuremath{2.8\pm0.6}}                     
\newcommand{\hatcurISOMK}{\ensuremath{2.59\pm0.11}}                    
\newcommand{\hatcurISOJK}{\ensuremath{0.29\pm0.02}}                    
\newcommand{\hatcurISOspec}{F8}                                        
\newcommand{\hatcurRVK}{\ensuremath{83.0\pm3.4}}                       
\newcommand{\hatcurRVk}{\ensuremath{-0.053\pm0.021}}                   
\newcommand{\hatcurRVh}{\ensuremath{-0.017\pm0.042}}                   
\newcommand{\hatcurRVgammaA}{\ensuremath{3.20\pm2.62}}                 
\newcommand{\hatcurRVeccen}{\ensuremath{0.067\pm0.024}}                
\newcommand{\hatcurRVomega}{\ensuremath{197\pm36}}                     
\newcommand{\hatcurPPi}{\ensuremath{88.6\pm0.7}}                       
\newcommand{\hatcurPPlogg}{\ensuremath{3.04\pm0.05}}                   
\newcommand{\hatcurPPar}{\ensuremath{7.58\pm0.35}}                     
\newcommand{\hatcurPParel}{\ensuremath{0.0465\pm0.0006}}               
\newcommand{\hatcurPPrho}{\ensuremath{0.44\pm0.07}}                    
\newcommand{\hatcurPPmlong}{\ensuremath{0.685\pm0.033}}                
\newcommand{\hatcurPPrlong}{\ensuremath{1.242\pm0.067}}                
\newcommand{\hatcurPPmrcorr}{\ensuremath{0.31}}                        
\newcommand{\hatcurPPteff}{\ensuremath{1637\pm42}}                     
\newcommand{\hatcurPPtheta}{\ensuremath{0.043\pm0.003}}                
\newcommand{\hatcurPPfluxperi}{\ensuremath{1.84\pm0.208}}              
\newcommand{\hatcurPPfluxperidim}{\ensuremath{9}}                      
\newcommand{\hatcurPPfluxap}{\ensuremath{1.44\pm0.168}}                
\newcommand{\hatcurPPfluxapdim}{\ensuremath{9}}                        
\newcommand{\hatcurPPfluxavg}{\ensuremath{1.62\pm0.169}}               
\newcommand{\hatcurPPfluxavgdim}{\ensuremath{9}}                       
\newcommand{\hatcurXsecondary}{\ensuremath{2455218.542\pm0.045}}       
\newcommand{\hatcurXsecdur}{\ensuremath{0.1492\pm0.0121}}              
\newcommand{\hatcurXsecingdur}{\ensuremath{0.0137\pm0.0013}}           
\newcommand{\hatcurXdist}{\ensuremath{396\pm20}}                       

\newcommand{\hatcur}{HAT-P-24}
\newcommand{\hatcurb}{HAT-P-24b}

\newcommand{\hatcurfieldtwo}{315}
\newcommand{\hatcurRVgammarel}{\hatcurRVgammaA}                           
\newcommand{\hatcurCCtassvi}{\ensuremath{0.628\pm0.089}}                  
\newcommand{\hatcurSMEversion}{ii}                                       
\newcommand{\hatcurSMEteff}{\ifthenelse{\equal{\hatcurSMEversion}{i}}{\hatcurSMEiteff}{\hatcurSMEiiteff}}
\newcommand{\hatcurSMEzfeh}{\ifthenelse{\equal{\hatcurSMEversion}{i}}{\hatcurSMEizfeh}{\hatcurSMEiizfeh}}
\newcommand{\hatcurSMEzfehshort}{\ifthenelse{\equal{\hatcurSMEversion}{i}}{\hatcurSMEizfehshort}{\hatcurSMEiizfehshort}}
\newcommand{\hatcurSMElogg}{\ifthenelse{\equal{\hatcurSMEversion}{i}}{\hatcurSMEilogg}{\hatcurSMEiilogg}}
\newcommand{\hatcurSMEvsin}{\ifthenelse{\equal{\hatcurSMEversion}{i}}{\hatcurSMEivsin}{\hatcurSMEiivsin}}
\newcommand{\hatcurSMEvmac}{\ifthenelse{\equal{\hatcurSMEversion}{i}}{\hatcurSMEivmac}{\hatcurSMEiivmac}}
\newcommand{\hatcurSMEvmic}{\ifthenelse{\equal{\hatcurSMEversion}{i}}{\hatcurSMEivmic}{\hatcurSMEiivmic}}

\newcommand{\hatcurisoshort}{YY}
\newcommand{\hatcurisocite}{yi:2001}
\newcommand{\hatcurlumind}{\arstar}
\newcommand{\hatcurjhkfilset}{ESO}

\newboolean{emulateapj}
\setboolean{emulateapj}{true}

\newboolean{rvtablelong}
\setboolean{rvtablelong}{true}

\newboolean{astroph}
\setboolean{astroph}{true}

\shortauthors{Kipping et al.}
\shorttitle{\hatcur\lowercase{b}}
\ifthenelse{\boolean{emulateapj}}{
    \newcommand{\titledag}{$\dagger$}
}{
    \newcommand{\titledag}{\dagger}
}

\begin{document}

\title{\hatcur\lowercase{b}: An inflated hot-Jupiter on a 3.36\lowercase{d}
	period transiting a hot, metal-poor star \altaffilmark{\titledag}}

\author{
   D.~M.~Kipping\altaffilmark{1,2},
   G.~\'A.~Bakos\altaffilmark{1,3},
   J.~Hartman\altaffilmark{1},
   G.~Torres\altaffilmark{1},
   A.~Shporer\altaffilmark{4,5}
   D.~W.~Latham\altaffilmark{1},
   G\'eza~Kov\'acs\altaffilmark{6},
   R.~W.~Noyes\altaffilmark{1},
   A.~W.~Howard\altaffilmark{7},
   D.~A.~Fischer\altaffilmark{8},
   J.~A.~Johnson\altaffilmark{9},
   G.~W.~Marcy\altaffilmark{7},
   B.~B\'eky\altaffilmark{1},
   G.~Perumpilly\altaffilmark{1}
   G.~A.~Esquerdo\altaffilmark{1},
   D.~D.~Sasselov\altaffilmark{1},
   R.~P.~Stefanik\altaffilmark{1},
   J.~L\'az\'ar\altaffilmark{10},
   I.~Papp\altaffilmark{10},
   P.~S\'ari\altaffilmark{10}
}
\altaffiltext{1}{Harvard-Smithsonian Center for Astrophysics,
	Cambridge, MA; email: dkipping@cfa.harvard.edu}

\altaffiltext{2}{University College London, Dept.~of Physics \&
                 Astronomy, Gower St., London, UK}

\altaffiltext{3}{NSF Fellow}

\altaffiltext{4}{Las Cumbres Observatory Global Telescope Network,
                 6740 Cortona Drive, Suite 102, Santa Barbara, CA 93106}

\altaffiltext{5}{Department of Physics, Broida Hall, University of 
                 California, Santa Barbara, CA 93106}

\altaffiltext{6}{Konkoly Observatory, Budapest, Hungary}

\altaffiltext{7}{Department of Astronomy, University of California,
	Berkeley, CA}

\altaffiltext{8}{Department of Astronomy, Yale University, New Haven}

\altaffiltext{9}{California Institute of Technology, Department of 
                 Astrophysics, MC 249-17, Pasadena, CA}

\altaffiltext{10}{Hungarian Astronomical Association, Budapest, 
	Hungary}

\altaffiltext{$\dagger$}{
	Based in part on observations obtained at the W.~M.~Keck
	Observatory, which is operated by the University of California and
	the California Institute of Technology. Keck time has been
	granted by NOAO and NASA\@.
}


\begin{abstract}

\setcounter{footnote}{10} 
We report the discovery of \hatcurb{}, a transiting extrasolar planet
orbiting the moderately bright V=\hatcurCCtassmv\ \hatcurISOspec\ dwarf
star \hatcurCCgsc, with a period $P=3.3552464 \pm 0.0000071$\,d,
transit epoch $T_c= 2455216.97669 \pm 0.00024$
(BJD\footnote{Barycentric Julian dates throughout the paper are
calculated from Coordinated Universal Time (UTC)}), and transit
duration $3.653 \pm 0.025$\,hours.
The host star has a mass of $1.191 \pm 0.042$\,\msun, radius of $1.317
\pm 0.068$\,\rsun, effective temperature \hatcurSMEteff\,K, and a low
metallicity of $\feh = \hatcurSMEzfeh$.
The planetary companion has a mass of $0.681 \pm 0.031$\,\mjup, and
radius of $1.243 \pm 0.072$\,\rjup\ yielding a mean density of $0.439
\pm 0.069$\,\gcmc.
By repeating our global fits with different parameter sets, we have
performed a critical investigation of the fitting techniques used for
previous HAT planetary discoveries.  We find that the system properties
are robust against the choice of priors.  The effects of fixed versus
fitted limb darkening are also examined.
HAT-P-24b probably maintains a small eccentricity of
$e=0.052_{-0.017}^{+0.022}$, which is accepted over the circular orbit
model with false alarm probability 5.8\%.  In the absence of
eccentricity pumping, this result suggests HAT-P-24b experiences less
tidal dissipation than Jupiter.
Due to relatively rapid stellar rotation, we estimate that HAT-P-24b
should exhibit one of the largest known Rossiter-McLaughlin effect
amplitudes for an exoplanet ($\Delta V_{RM} \simeq 95$\,m/s) and thus a
precise measurement of the sky-projected spin-orbit alignment should be
possible.

\setcounter{footnote}{0}
\end{abstract}



\section{Introduction}
\label{sec:introduction}

The understanding of planetary systems has been spearheaded by the
study of transiting extrasolar planets in recent years.  Eclipses have
long offered a key to unlocking the secrets of the heavenly bodies, for
example in Solar System studies and in the field of eclipsing binaries. 
For exoplanets, an eclipse offers a door into the inner-workings of an
alien system hundreds of light-years away.  As an inherently low
probability event, each and every transiting system is precious and
timeless to the planetary scientist.  Transits continue to offer 
unprecedented access to an exoplanet's nature and allow for, amongst
other things, the determination of the oblateness of a planet 
(\citet{seager:2002}; \citet{carter:2010}), thermal mapping of the 
planetary surface \citep{knutson:2007} and accurate planetary radii
at the percent level \citep{dc:2000}.

The Hungarian-made Automated Telescope Network
\citep[HATNet;][]{bakos:2004} survey has been one of the principal
contributors to the discovery of transiting extrasolar planets (TEPs). 
In operation since 2003, it has now covered approximately 14\% of the
sky, searching for TEPs around bright stars ($8\lesssim I \lesssim
14.0$).  HATNet operates six wide-field instruments: four at the Fred
Lawrence Whipple Observatory (FLWO) in Arizona, and two on the roof of
the hangar housing the Smithsonian Astrophysical Observatory's
Submillimeter Array, in Hawaii. Since 2006, HATNet has discovered 23
TEPs (with 16 announced or published so far).  In this work we report
our 24th discovery, around the relatively bright star previously known
as \hatcurCCgsc{}.

In \refsecl{obs} we report the
detection of the photometric signal and the follow-up spectroscopic and
photometric observations of \hatcur{}.  In \refsecl{analysis} we
describe the analysis of the data, beginning with the determination of
the stellar parameters, continuing with a discussion of the methods
used to rule out non-planetary, false positive scenarios which could
mimic the photometric and spectroscopic observations, and finishing
with a description of our global modeling of the photometry and radial
velocities.  In \refsecl{priors} we investigate the variations of our
results using different parameter sets with uniform priors to test the
robustness of the fitted parameters.  We also discuss the effects of
fitting versus fixing limb darkening coefficients.  In \refsecl{ecc} 
to \refsecl{followup}, we present discussions and analyses of the 
orbital eccentricity, a possible linear drift in the RVs, the measured
mid-transit times and possibilities for future follow-up observations. 
Finally, we summarize our findings in \refsecl{summary}.

\section{Observations}
\label{sec:obs}

\subsection{Photometric detection}
\label{sec:detection}

The transits of \hatcurb{} were detected with the HAT-5 and HAT-6
telescopes in Arizona, and with the HAT-8 and HAT-9 telescopes in
Hawaii.  The star \hatcurCCgsc{} lies in the overlap of two fields,
internally labeled as \hatcurfieldtwo{} (07:30 +15:00) and 
\hatcurfield{} (07:00 +15:00).  The former field was observed on a 
nightly basis between 2007 October and 2008 May, while the latter 
field was observed between 2008 November and 2009 May.  For field 
\hatcurfieldtwo{} we gathered 8551 \band{R} exposures of 5 minutes at 
a 5.5 minute cadence, while for field \hatcurfield{} we gathered 5503 
Sloan \band{r} images with the same exposure time and cadence.  
Each field \hatcurfieldtwo{} image contained approximately
51,000 stars down to $R\sim14$, while each \hatcurfield{} image
contained approximately 130,000 stars down to $r\sim14.5$.  For the
brightest stars in field \hatcurfieldtwo, we achieved a per-image
photometric precision of 3\,mmag, while for field \hatcurfield{} we
achieved 5\,mmag precision.

\begin{figure}[!ht]
\plotone{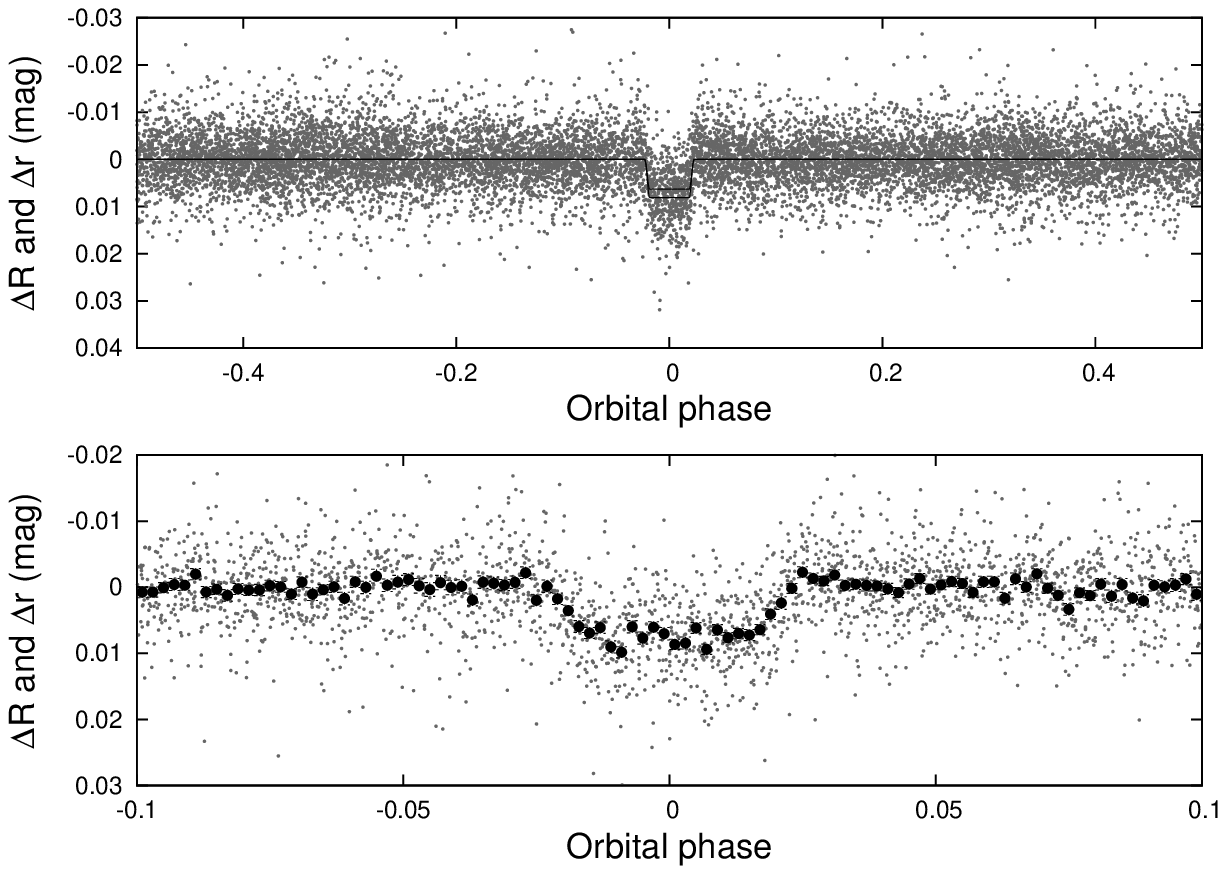}
\caption{
	Unbinned \lc{} of \hatcur{} including all 14,000 instrumental
    \band{R} and Sloan \band{r} 5.5 minute cadence measurements
    obtained with the HAT-5, HAT-6, HAT-8 and HAT-9 telescopes of
    HATNet (see the text for details), and folded with the period $P =
    \hatcurLCPprec$\,days resulting from the global fit described in
    \refsecl{analysis}).  The solid line shows the ``P1P3'' transit
    model fit to the light curve (\refsecl{globmod}).  Solid squares
    show the 20-point binned light curve.
\label{fig:hatnet}}
\end{figure}

The calibration of the HATNet frames was carried out using standard
photometric procedures.  For field \hatcurfield{} the calibrated images
were then subjected to star detection and astrometry, as described in
\cite{pal:2006}.  Aperture photometry was performed on each image at
the stellar centroids derived from the Two Micron All Sky Survey
\citep[2MASS;][]{skrutskie:2006} catalog and the individual astrometric
solutions.  For field \hatcurfieldtwo{} we performed image subtraction
photometry following the methods described in \cite{pal:2009b}. 
Aperture photometry was performed on the difference images at the
stellar centroid derived from 2MASS, while for the reference flux we
adopted the $r$ magnitude of each star, transformed from its 2MASS $J$,
$H$ and $K_S$ magnitudes, and made use of the average relation between
$r$ and flux measured on the reference image via aperture photometry. 
The resulting \lcs\ for both fields were decorrelated (cleaned of
trends) using the External Parameter Decorrelation \citep[EPD;
see][]{bakos:2009} technique in ``constant'' mode and the Trend
Filtering Algorithm \citep[TFA; see][]{kovacs:2005}.  The \lcs{} from
each field were independently searched for periodic box-shaped signals
using the Box Least-Squares \citep[BLS; see][]{kovacs:2002} method.  We
detected a significant signal in both of the HATNet \lcs{} of
\hatcurCCgsc{} (also known as \hatcurCCtwomass{}; $\alpha =
\hatcurCCra$, $\delta = \hatcurCCdec$; J2000; V=\hatcurCCtassmv{}
\citealp{droege:2006}), with an apparent depth of
$\sim\hatcurLCdip$\,mmag, and a period of $P=\hatcurLCPshort$\,days
(see \reffigl{hatnet}).  

\subsection{Reconnaissance Spectroscopy}
\label{sec:recspec}

As is routine in the HATNet project, all candidates are observed
spectroscopically initially to establish
whether the transit-like feature in the light curve is of non-planetary
origin such a grazing eclipsing binary (i.e.~a false positive).  
For example, large radial-velocity
variations of the star (tens of \kms) would indicate such a
circumstance.

To perform this task, we used the Harvard-Smithsonian Center for
Astrophysics (CfA) Digital Speedometer \citep[DS;][]{latham:1992}; an
echelle spectrograph mounted on the \flwos\ telescope.  This instrument
delivers high-resolution spectra ($\lambda/\Delta\lambda \approx
35,\!000$) over a single order centered on the \ion{Mg}{1}\,b triplet
($\sim$5187\,\AA), with typically low signal-to-noise (S/N$\sim10$) ratios that
are nevertheless sufficient to derive radial velocities (RVs) with
moderate precisions of 0.5--1.0\,\kms\ for slowly rotating stars.  The
same spectra can be used to estimate the effective temperature, surface
gravity, and projected rotational velocity of the host star, as
described by \cite{torres:2002}.  With this facility we are able to
reject many types of false positives, such as F dwarfs orbited by M
dwarfs, grazing eclipsing binaries, or triple or quadruple star
systems.  

For \hatcur{} we obtained five observations with the DS between 2008
May and 2009 January.  The velocity measurements showed an r.m.s.~
residual of \hatcurDSrvrms\,\kms, consistent with no detectable RV
variation within the precision of the measurements.  All spectra were
single-lined, i.e., there is no evidence for additional stars in the
system.  The atmospheric parameters we infer from these observations
are the following: effective temperature $\teffstar =
\hatcurDSteff\,K$, surface gravity $\loggstar = \hatcurDSlogg$ (log
cgs), and projected rotational velocity $\vsini =
\hatcurDSvsini\,\kms$.  The effective temperature corresponds to a
\hatcurISOspec\ dwarf.  The mean heliocentric RV of \hatcur\ is
$\gamma_{\rm RV} = -2.09 \pm 0.33$\,\kms.  We stress that the DS
stellar parameters are based on solar composition models.

\subsection{High resolution, high S/N spectroscopy}
\label{sec:hispec}

Given the significant transit detection by HATNet, and the encouraging
DS results that rule out obvious false positives, we proceeded with the
follow-up of this candidate by obtaining high-resolution, high-S/N
spectra to characterize the RV variations, and to refine the
determination of the stellar parameters.  For this we used the HIRES
instrument \citep{vogt:1994} on the Keck~I telescope located on Mauna
Kea, Hawaii, between 2009 April and 2009 December.  The width of the
spectrometer slit was $0\farcs86$, resulting in a resolving power of
$\lambda/\Delta\lambda \approx 55,\!000$, with a wavelength coverage of
$\sim$3800--8000\,\AA\@.

\begin{figure}[ht]
\ifthenelse{\boolean{emulateapj}}{
	\plotone{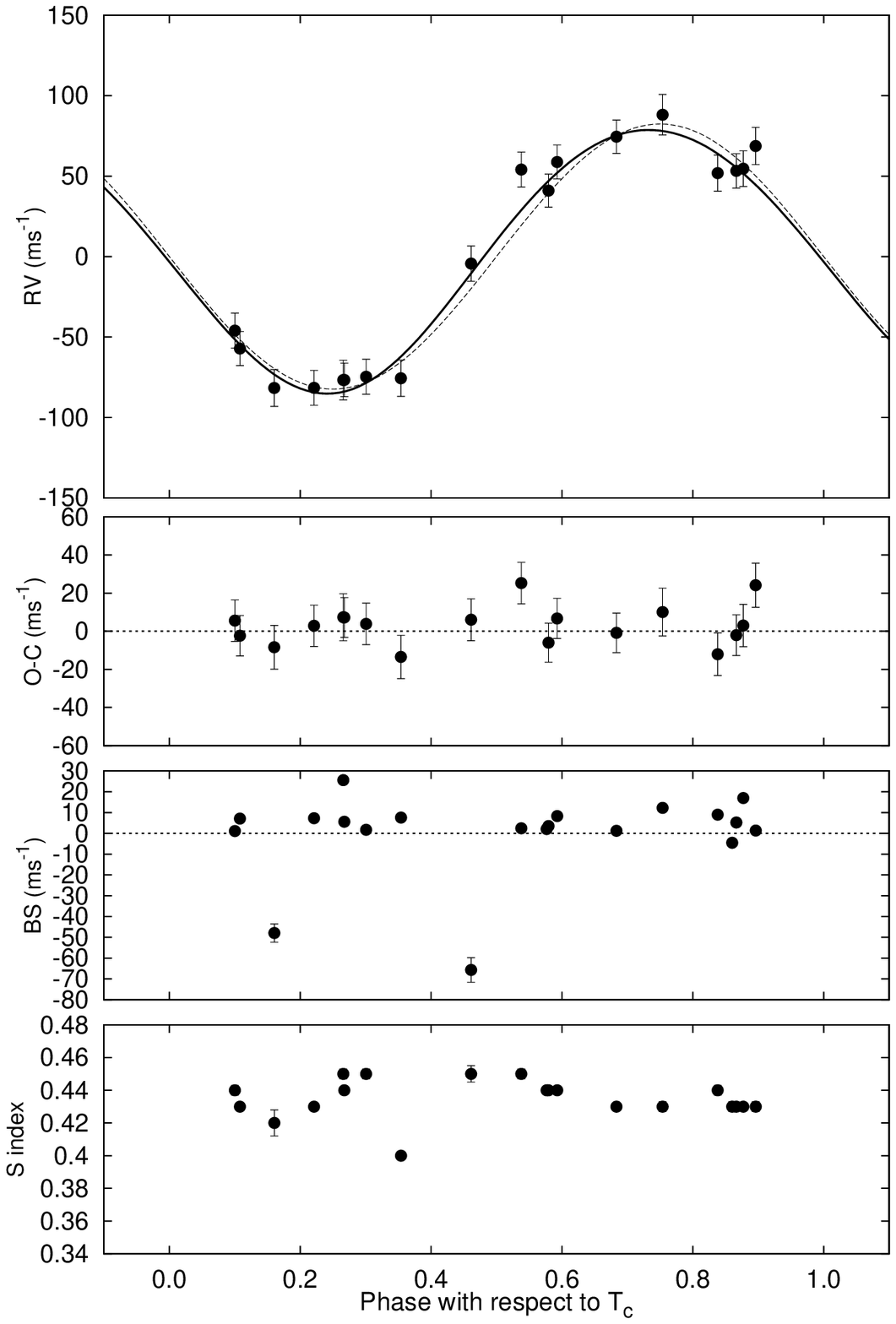}
}{
	\includegraphics[scale=0.8]{\hatcurhtr-rv.eps}
}
\caption{
	{\em Top panel:} Keck/HIRES RV measurements for
    \hbox{\hatcur{}} shown as a function of orbital phase, along with
    our best-fit eccentric orbit model (solid) and circular orbit 
    model (dashed).  Zero phase
    corresponds to the time of mid-transit.  The center-of-mass
    velocity has been subtracted.
	{\em Second panel:} Phase folded velocity $O\!-\!C$ residuals from
    the best fit.  The error bars include a component from
    astrophysical jitter ($7.4$\,\ms) added in quadrature to the formal
    errors (see \refsecl{globmod}).
	{\em Third panel:} Bisector spans (BS), with the mean value
    subtracted.  The measurement from the template spectrum is included
    (see \refsecl{bisec}).
	{\em Bottom panel:} Relative chromospheric activity index $S$
    measured from the Keck spectra.
	Note the different vertical scales of the panels. 
\label{fig:rvbis}}
\end{figure}

We obtained 18 exposures through an iodine gas absorption cell, which
was used to superimpose a dense forest of $\mathrm{I}_2$ lines on the
stellar spectrum and establish an accurate wavelength fiducial
\citep[see][]{marcy:1992}.  An additional two exposures were taken
without the iodine cell, for use as templates in the reductions.  In
practice, we used only the second higher S/N template spectrum. 
Relative RVs in the solar system barycentric frame were derived as
described by \cite{butler:1996}, incorporating full modeling of the
spatial and temporal variations of the instrumental profile.  The RV
measurements and their uncertainties are listed in \reftabl{rvs}.  The
period-folded data, along with a best fit described below in
\refsecl{analysis}, are displayed in \reffigl{rvbis}.

In the same figure we show also the relative $S$ index, which is a
measure of the chromospheric activity of the star derived from the flux
in the cores of the \ion{Ca}{2} H and K lines.  This index was computed
following the prescription given by \citet{vaughan:1978}, after
matching each spectrum to a reference spectrum using a transformation
that includes a wavelength shift and a flux scaling that is a
polynomial as a function of wavelength.  The transformation was
determined on regions of the spectra that are not used in computing
this indicator. Note that our relative $S$ index has not been calibrated to
the scale of \citet{vaughan:1978}.
The RMS of the relative S values is 3\%, which is higher than the median
formal error of 0.6\% based on photon statistics, however the errors in this
case are likely dominated by systematics in the spectrum matching 
procedure, which are difficult to quantify, so we do not consider this to be
a robust detection of variability. We note that a 3\% variation is comparable
to that found for other late F stars (e.g. \citealp{shkolnik:2008} measured a 
 $\sim 1\%$ median absolute deviation in the fluxes of the K line cores 
 for $\tau$~Boo and HD~179949, which corresponds to a similar 
 expected RMS).

We also note that $S$ is uncorrelated with orbital phase; 
such a correlation might have
indicated that the RV variations could be due to stellar activity,
casting doubt on the planetary nature of the candidate.
There is no sign of emission in the cores of the \ion{Ca}{2} H and K
lines (S/N$\sim38$) in any of our spectra, from which we conclude that 
the chromospheric activity level in \hatcur{} is very low.

\ifthenelse{\boolean{emulateapj}}{
    \begin{deluxetable*}{lrrrrrr}	
}{
    \begin{deluxetable}{lrrrrrr}
	\tabletypesize{\scriptsize}
}
\tablewidth{0pc}
\tablecaption{
	Relative radial velocities, bisector spans, and activity index
	measurements of \hatcur{}.
	\label{tab:rvs}
}
\tablehead{
	\colhead{BJD} & 
	\colhead{RV\tablenotemark{a}} & 
	\colhead{\ensuremath{\sigma_{\rm RV}}\tablenotemark{b}} & 
	\colhead{BS} & 
	\colhead{\ensuremath{\sigma_{\rm BS}}} & 
	\colhead{S\tablenotemark{c}} &
	\colhead{\ensuremath{\sigma_{\rm S}}\tablenotemark{c}} \\
	\colhead{\hbox{(2,454,000$+$)}} & 
	\colhead{(\ms)} & 
	\colhead{(\ms)} &
	\colhead{(\ms)} &
    \colhead{(\ms)} &
	\colhead{} &
	\colhead{}
}
\startdata
\ifthenelse{\boolean{rvtablelong}}{
    $ 928.76217 $ \dotfill & $   -49.49 $ & $     6.43 $ & $     1.10 $ & $     1.03 $ & $    0.44 $ & 0.003 \\
$ 954.79832 $ \dotfill & \nodata      & \nodata      & $    -4.52 $ & $     1.35 $ & $    0.43 $ & 0.003 \\
$ 955.80587 $ \dotfill & $   -85.15 $ & $     7.37 $ & $   -47.96 $ & $     4.39 $ & $    0.42 $ & 0.008 \\
$ 956.81494 $ \dotfill & $    -7.84 $ & $     6.61 $ & $   -65.72 $ & $     5.91 $ & $    0.45 $ & 0.005 \\
$ 963.78272 $ \dotfill & $    50.62 $ & $     6.41 $ & $     2.42 $ & $     0.68 $ & $    0.45 $ & 0.003 \\
$ 1107.14583 $ \dotfill & $   -80.20 $ & $     8.64 $ & $    25.58 $ & $     1.21 $ & $    0.45 $ & 0.003 \\
$ 1109.06567 $ \dotfill & $    48.40 $ & $     7.00 $ & $     8.95 $ & $     0.13 $ & $    0.44 $ & 0.003 \\
$ 1112.13808 $ \dotfill & $    84.68 $ & $     9.02 $ & $    12.27 $ & $     0.08 $ & $    0.43 $ & 0.003 \\
$ 1134.10472 $ \dotfill & $   -78.19 $ & $     6.49 $ & $     1.68 $ & $     0.92 $ & $    0.45 $ & 0.003 \\
$ 1135.08419 $ \dotfill & $    55.31 $ & $     5.83 $ & $     8.32 $ & $     0.41 $ & $    0.44 $ & 0.002 \\
$ 1136.10198 $ \dotfill & $    65.26 $ & $     7.53 $ & $     1.35 $ & $     0.77 $ & $    0.43 $ & 0.002 \\
$ 1172.94629 $ \dotfill & $    51.15 $ & $     6.80 $ & $    16.94 $ & $     0.21 $ & $    0.43 $ & 0.002 \\
$ 1174.10042 $ \dotfill & $   -85.04 $ & $     6.32 $ & $     7.30 $ & $     0.48 $ & $    0.43 $ & 0.001 \\
$ 1187.96722 $ \dotfill & $   -79.11 $ & $     7.17 $ & $     7.57 $ & $     0.35 $ & $    0.40 $ & 0.002 \\
$ 1189.07232 $ \dotfill & $    71.01 $ & $     5.53 $ & $     1.20 $ & $     0.80 $ & $    0.43 $ & 0.001 \\
$ 1191.03182 $ \dotfill & $   -80.15 $ & $     5.68 $ & $     5.56 $ & $     0.56 $ & $    0.44 $ & 0.002 \\
$ 1192.06936 $ \dotfill & \nodata      & \nodata      & $     2.08 $ & $     0.77 $ & $    0.44 $ & 0.001 \\
$ 1192.07902 $ \dotfill & $    37.54 $ & $     5.33 $ & $     3.51 $ & $     0.66 $ & $    0.44 $ & 0.001 \\
$ 1193.04199 $ \dotfill & $    49.79 $ & $     6.10 $ & $     5.20 $ & $     0.59 $ & $    0.43 $ & 0.002 \\
$ 1193.85228 $ \dotfill & $   -60.65 $ & $     5.88 $ & $     7.09 $ & $     0.52 $ & $    0.43 $ & 0.002 \\

	[-1.5ex]
}{
    $ 928.76217 $ \dotfill & $   -49.49 $ & $     6.43 $ & $     1.10 $ & $     1.03 $ & $    0.44 $ \\
$ 954.79832 $ \dotfill & \nodata      & \nodata      & $    -4.52 $ & $     1.35 $ & $    0.43 $ \\
$ 955.80587 $ \dotfill & $   -85.15 $ & $     7.37 $ & $   -47.96 $ & $     4.39 $ & $    0.42 $ \\

	[-1.5ex]
}
\enddata
\tablenotetext{a}{
	The zero-point of these velocities is arbitrary. An overall
    offset $\gamma_{\rm rel}$ fitted to these velocities in
    \refsecl{globmod} has {\em not} been subtracted.
}
\tablenotetext{b}{
	Internal errors excluding the component of astrophysical jitter
    considered in \refsecl{globmod}.
}
\tablenotetext{c}{
	Relative chromospheric activity index, not calibrated to the
	scale of \citet{vaughan:1978}.
}
\ifthenelse{\boolean{rvtablelong}}{
	\tablecomments{
		Note that for the iodine-free template exposures we do not
		measure the RV but do measure the BS and S index.  Such
		template exposures can be distinguished by the missing RV
		value.
	}
}{
    \tablecomments{
		Note that for the iodine-free template exposures we do not
		measure the RV but do measure the BS and S index.  Such
		template exposures can be distinguished by the missing RV
		value.  This table is presented in its entirety in the
		electronic edition of the Astrophysical Journal.  A portion is
		shown here for guidance regarding its form and content.
	}
} 
\ifthenelse{\boolean{emulateapj}}{
    \end{deluxetable*}
}{
    \end{deluxetable}
	\clearpage
}

\subsection{Photometric follow-up observations}
\label{sec:phot}

\begin{figure}[!ht]
\plotone{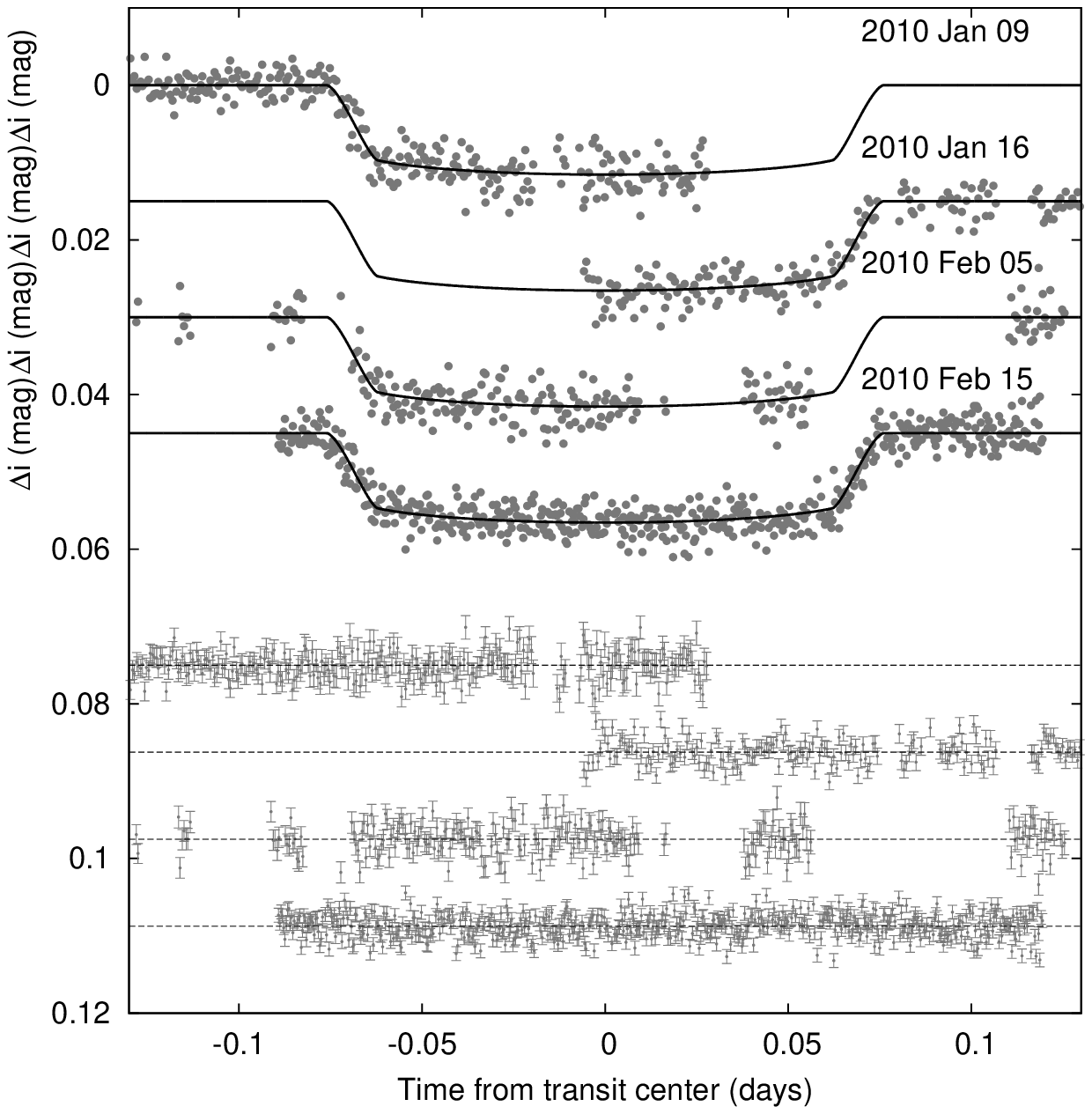}
\caption{
	Unbinned normalized Sloan \band{i} transit \lcs{}, acquired with
    KeplerCam at the \flwof{} telescope on 2010 Jan 09, 2010 Jan 16,
    and 2010 Feb 05, and with the Faulkes Telescope North on 2010 Feb
    15.  The light curves have been EPD and TFA processed, as described
    in \refsec{globmod}.  Final fits come from the
    \{$p^2$,$b$,$T_{1.5,3.5}$\} light curve parameter set and fitted
    limb darkening, in combination with an eccentric orbit RV fit.
    The dates of the events are indicated.  Curves after the first are
    displaced vertically for clarity.  Our best fit from the global
    modeling described in \refsecl{globmod} is shown by the solid
    lines.  Residuals from the fits are displayed at the bottom, in the
    same order as the top curves.  The error bars represent the photon
    and background shot noise, plus the readout noise.
\label{fig:lc}}
\end{figure}

In order to permit a more accurate modeling of the light curve, we
conducted additional photometric observations with the KeplerCam CCD
camera on the \flwof{} telescope in Arizona and with the 2.0\,m Faulkes
Telescope North (FTN) at Haleakala Observatory in Hawaii.  We observed
three transit events of \hatcur{} with the \flwof{} telescope on the
nights of 2010 Jan 09, 2010 Jan 16 and 2010 Feb 05, and a fourth
transit event with the FTN on the night of 2010 Feb 15 (\reffigl{lc}). 
These observations are summarized in \reftabl{phfusummary}.
%

\ifthenelse{\boolean{emulateapj}}{
    \begin{deluxetable*}{llrrr}
}{
    \begin{deluxetable}{llrr}
	\tabletypesize{\scriptsize}
}
\tablewidth{0pc}
\tabletypesize{\scriptsize}
\tablecaption{
    Summary of photometric follow-up observations, all of
    which were taken in Sloan \band{i}.
    \label{tab:phfusummary}
}
\tablehead{
    \colhead{Facility}  &
    \colhead{Date} &
    \colhead{Number of Images} &
    \colhead{Cadence (s)}
}
\startdata
KeplerCam/\flwof{} & 2010 Jan 09 & 527 & 39 \\
KeplerCam/\flwof{} & 2010 Jan 16 & 246 & 53 \\
KeplerCam/\flwof{} & 2010 Feb 05 & 256 & 39 \\
FTN                & 2010 Feb 15 & 534 & 30 \\
[-1.5ex]
\enddata
\ifthenelse{\boolean{emulateapj}}{
    \end{deluxetable*}
}{
    \end{deluxetable}
}

%
The reduction of these images, including basic calibration, astrometry,
and aperture photometry, was performed as described by
\citet{bakos:2009}.  We performed EPD and TFA to remove trends
simultaneously with the light curve modeling (for more details, see
\refsecl{analysis}, and \citealt{bakos:2009}).  The final time series
are shown in the top portion of \reffigl{lc}, along with our best-fit
transit \lc{} model described below; the individual measurements are
reported in \reftabl{phfu}.

\begin{deluxetable}{lrrr}
\tablewidth{0pc}
\tablecaption{High-precision differential photometry of 
	\hatcur\label{tab:phfu}}
\tablehead{
	\colhead{BJD} & 
	\colhead{Mag\tablenotemark{a}} & 
	\colhead{\ensuremath{\sigma_{\rm Mag}}} &
	\colhead{Filter} \\
	\colhead{\hbox{~~~~(2,400,000$+$)~~~~}} & 
	\colhead{} & 
	\colhead{} &
	\colhead{}
}
\startdata
$ 55206.63289 $ & $  -0.00145 $ & $   0.00176 $ & $ i$\\
$ 55206.63724 $ & $   0.00220 $ & $   0.00181 $ & $ i$\\
$ 55206.63790 $ & $  -0.00278 $ & $   0.00173 $ & $ i$\\
$ 55206.64440 $ & $  -0.00213 $ & $   0.00174 $ & $ i$\\
$ 55206.64485 $ & $   0.00093 $ & $   0.00171 $ & $ i$\\
$ 55206.64548 $ & $  -0.00051 $ & $   0.00161 $ & $ i$\\
$ 55206.64594 $ & $   0.00535 $ & $   0.00161 $ & $ i$\\
$ 55206.64659 $ & $  -0.00391 $ & $   0.00141 $ & $ i$\\
$ 55206.64703 $ & $   0.00162 $ & $   0.00140 $ & $ i$\\
$ 55206.64767 $ & $  -0.00023 $ & $   0.00134 $ & $ i$\\

[-1.5ex]
\enddata
\tablenotetext{a}{
	The out-of-transit level has been subtracted. These magnitudes have
	been subjected to the EPD and TFA procedures, carried out
	simultaneously with the transit fit.
}
\tablecomments{
    The complete table is available in a machine-readable form in the on-line
    journal.  A portion is shown here for guidance regarding its form
    and content.
}
\end{deluxetable}

\section{Analysis}
\label{sec:analysis}

\subsection{Properties of the parent star}
\label{sec:stelparam}

Fundamental parameters of the host star \hatcur{} such as the mass
(\mstar) and radius (\rstar), which are needed to infer the planetary
properties, depend strongly on other stellar quantities that can be
derived spectroscopically.  For this we have relied on our template
spectrum obtained with the Keck/HIRES instrument, and the analysis
package known as Spectroscopy Made Easy \citep[SME;][]{valenti:1996},
along with the atomic line database of \cite{valenti:2005}.  SME
yielded the following {\em initial} values and uncertainties (which we
have conservatively increased for \teffstar\ and \feh\ to include our
estimates of the systematic errors):
effective temperature $\teffstar=\hatcurSMEiteff$\,K, stellar surface
gravity $\loggstar=\hatcurSMEilogg$\,(cgs), metallicity
$\feh=\hatcurSMEizfeh$\,dex, and projected rotational velocity
$\vsini=\hatcurSMEivsin\,\kms$.

In principle the effective temperature and metallicity, along with the
surface gravity taken as a luminosity indicator, could be used as
constraints to infer the stellar mass and radius by comparison with
stellar evolution models.
For planetary transits a stronger constraint is often provided by the
\arstar\ normalized semi-major axis, which is closely related to
\rhostar, the mean stellar density.  The quantity \arstar\ can be
derived directly from the transit \lcs\ \citep{seager:2003} and the RV
data (for eccentric cases, see \citet{kipping:2010}).  This, in turn,
allows us to improve on the determination of the spectroscopic
parameters by supplying an indirect constraint on the weakly determined
spectroscopic value of \loggstar\ that removes degeneracies.  We take
this approach here, as described in \citet{bakos:2009}.  The validity
of our assumption, namely that the adequate physical model describing
our data is a planetary transit (as opposed to a blend), is shown later
in \refsecl{bisec}.

After the first iteration for determining the stellar properties, as
described in \citet{bakos:2009}, we find that the surface gravity,
$\loggstar = \hatcurISOlogg$, is significantly different from our
initial SME analysis, which is not surprising in view of the strong
correlations among \teffstar, \feh, and \loggstar\ that are often
present in spectroscopic determinations.  Therefore, we carried out a
second iteration in which we adopted this value of \loggstar\ and held
it fixed in a new SME analysis (coupled with a new global modeling of
the RV and \lcs), adjusting only \teffstar, \feh, and \vsini.  This
gave
$\teffstar = \hatcurSMEiiteff$\,K, 
$\feh = \hatcurSMEiizfeh$, and 
$\vsini = \hatcurSMEiivsin$\,\kms,
in which the conservative uncertainties for the first two have been
increased by a factor of two over their formal values, as before.
Experience with the SME analysis for previous HAT planets leads us
to select this estimate.  
A further iteration did not change \loggstar\ significantly, so we
adopted the values stated above as the final atmospheric properties of
the star.  They are collected in \reftabl{stellar}, together with the
adopted values for the macroturbulent and microturbulent velocities.

The low metallicity of HAT-P-24 is interesting in that a well-known
bias exists for finding giant planets around metal-rich stars 
\citep{johnson:2010}. \citet{johnson:2010} find that the occurrence
of giant planets scales as $f \sim 10^{1.2 \mathrm{[Fe/H]}}$
and also report a scaling with stellar mass of $f \sim M_*$.
This means that the a-priori probability of finding a planet around
HAT-P-24 is a respectable $\sim 75$\% that of a Solar-like star.

\begin{figure}[!ht]
\plotone{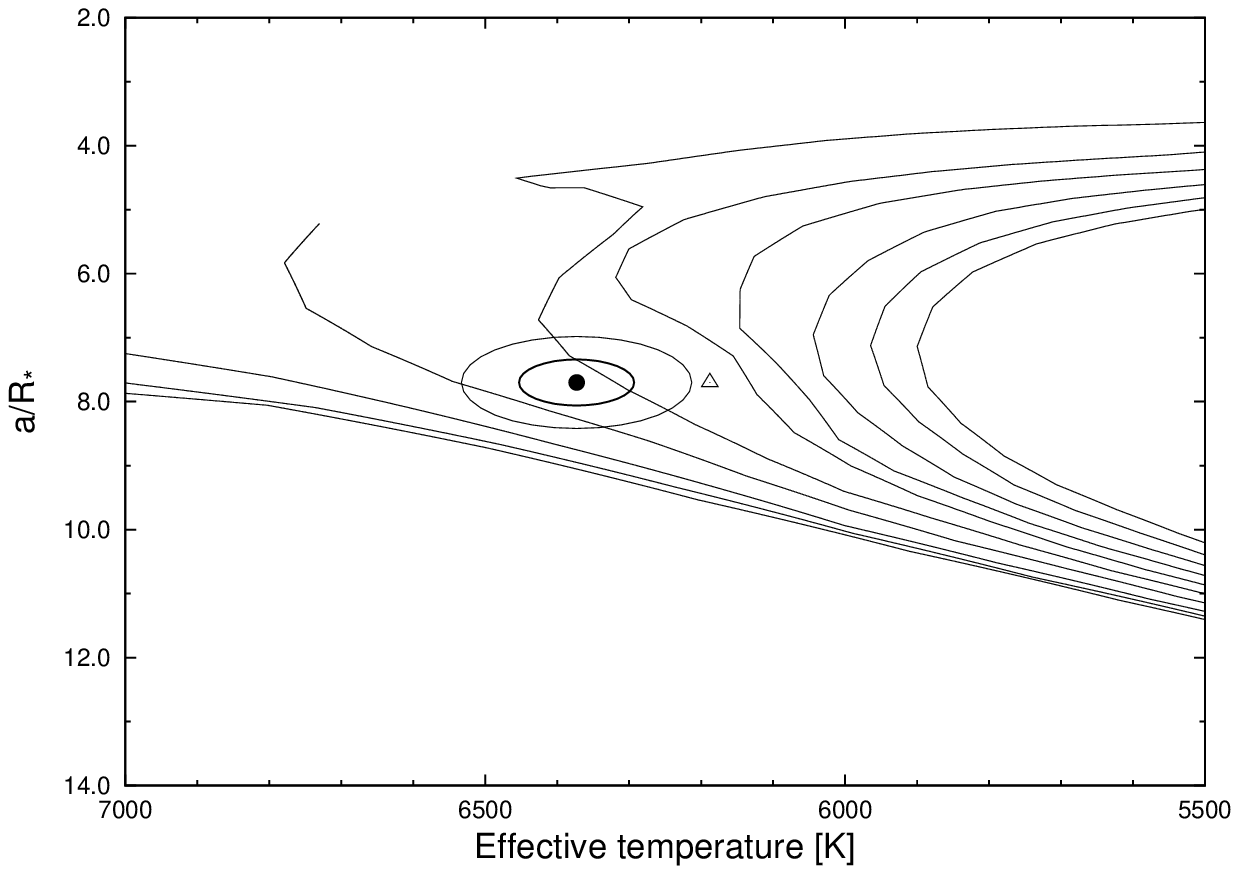}
\caption{
	Model isochrones from \cite{\hatcurisocite} for the measured
    metallicity of \hatcur, \feh = \hatcurSMEiizfehshort, and ages of
    0.2, 0.5, 1.0, 2.0, 3.0, 4.0, 5.0, 6.0, 7.0 and 8.0\,Gyr (left to
    right).  The adopted values of $\teffstar$ and \arstar\ are shown
    together with their 1$\sigma$ and 2$\sigma$ confidence ellipsoids. 
    The results come from the \{$p^2$,$b$,$T_{1.5,3.5}$\} light curve
    parameter set and fitted limb darkening, in combination with an
    eccentric orbit RV fit.  The initial values of \teffstar\ and
    \arstar\ from the first SME and \lc\ analyses are represented with
    a triangle.
\label{fig:iso}}
\end{figure}

With the adopted spectroscopic parameters the model isochrones yield
the stellar mass and radius \mstar\ = \hatcurISOmlong\,\msun\ and
\rstar\ = \hatcurISOrlong\,\rsun, along with other properties listed at
the bottom of \reftabl{stellar}.  \hatcur{} is a \hatcurISOspec\ dwarf
star with an estimated age of \hatcurISOage\,Gyr, according to these
models \citep{yi:2001}.  
The inferred location of the star in a diagram of \arstar\
versus \teffstar, analogous to the classical H-R diagram, is shown in
\reffigl{iso}.  The stellar properties and their 1$\sigma$ and
2$\sigma$ confidence ellipsoids are displayed against the backdrop of
\cite{\hatcurisocite} isochrones for the measured metallicity of \feh\
= \hatcurSMEiizfehshort, and a range of ages.  For comparison, the
location implied by the initial SME results is also shown (triangle),
and corresponds to a somewhat more evolved state.

The stellar evolution modeling provides color indices that may be
compared against the measured values as a consistency check.  The best
available measurements are the near-infrared magnitudes from the 2MASS
Catalogue \citep{skrutskie:2006},
$J_{\rm 2MASS}=\hatcurCCtwomassJmag$, 
$H_{\rm 2MASS}=\hatcurCCtwomassHmag$ and 
$K_{\rm 2MASS}=\hatcurCCtwomassKmag$;
which we have converted to the photometric system of the models (ESO
system) using the transformations by \citep{carpenter:2001}.  The
resulting measured color index is $J-K = \hatcurCCesoJKmag$.  This is
within 1$\sigma$ of the predicted value from the isochrones of $J-K =
\hatcurISOJK$.  The distance to the object may be computed from the
absolute $K$ magnitude from the models ($M_{\rm K}=\hatcurISOMK$) and
the 2MASS $K_s$ magnitude, which has the advantage of being less
affected by extinction than optical magnitudes.  The result is
$\hatcurXdist$\,pc, where the uncertainty excludes possible systematics
in the model isochrones that are difficult to quantify.

\begin{deluxetable}{lrl}
\tablewidth{0pc}
\tabletypesize{\scriptsize}
\tablecaption{
	Stellar parameters for \hatcur{}
	\label{tab:stellar}
}
\tablehead{
	\colhead{~~~~~~~~Parameter~~~~~~~~}	&
	\colhead{Value}                         &
	\colhead{Source}
}
\startdata
\noalign{\vskip -3pt}
\sidehead{Spectroscopic properties}
~~~~$\teffstar$ (K)\dotfill         &  \hatcurSMEteff   & SME\tablenotemark{a}\\
~~~~$\feh$\dotfill                  &  \hatcurSMEzfeh   & SME                 \\
~~~~$\vsini$ (\kms)\dotfill         &  \hatcurSMEvsin   & SME                 \\
~~~~$\vmac$ (\kms)\dotfill          &  \hatcurSMEvmac   & SME                 \\
~~~~$\vmic$ (\kms)\dotfill          &  \hatcurSMEvmic   & SME                 \\
~~~~$\gamma_{\rm RV}$ (\kms)\dotfill&  \hatcurDSgamma   & DS                  \\
\sidehead{Photometric properties}
~~~~$V$ (mag)\dotfill               &  \hatcurCCtassmv  & TASS                \\
~~~~$V\!-\!I_C$ (mag)\dotfill       &  \hatcurCCtassvi  & TASS                \\
~~~~$J$ (mag)\dotfill               &  \hatcurCCtwomassJmag & 2MASS           \\
~~~~$H$ (mag)\dotfill               &  \hatcurCCtwomassHmag & 2MASS           \\
~~~~$K_s$ (mag)\dotfill             &  \hatcurCCtwomassKmag & 2MASS           \\
\sidehead{Derived properties}
~~~~$\mstar$ ($\msun$)\dotfill      &  \hatcurISOmlong   & \hatcurisoshort+\hatcurlumind+SME \tablenotemark{b}\\
~~~~$\rstar$ ($\rsun$)\dotfill      &  \hatcurISOrlong   & \hatcurisoshort+\hatcurlumind+SME         \\
~~~~$\loggstar$ (cgs)\dotfill       &  \hatcurISOlogg    & \hatcurisoshort+\hatcurlumind+SME         \\
~~~~$\lstar$ ($\lsun$)\dotfill      &  \hatcurISOlum     & \hatcurisoshort+\hatcurlumind+SME         \\
~~~~$M_V$ (mag)\dotfill             &  \hatcurISOmv      & \hatcurisoshort+\hatcurlumind+SME         \\
~~~~$M_K$ (mag,\hatcurjhkfilset)\dotfill &  \hatcurISOMK & \hatcurisoshort+\hatcurlumind+SME         \\
~~~~Age (Gyr)\dotfill               &  \hatcurISOage     & \hatcurisoshort+\hatcurlumind+SME         \\
~~~~Distance (pc)\dotfill           &  \hatcurXdist      & \hatcurisoshort+\hatcurlumind+SME         \\
[-1.5ex]
\enddata
\tablenotetext{a}{
	SME = ``Spectroscopy Made Easy'' package for the analysis of
	high-resolution spectra \citep{valenti:1996}.  These parameters
	rely primarily on SME, but have a small dependence also on the
	iterative analysis incorporating the isochrone search and global
	modeling of the data, as described in the text.
}
\tablenotetext{b}{
	\hatcurisoshort+\hatcurlumind+SME = Based on the \hatcurisoshort\
    isochrones \citep{\hatcurisocite}, \hatcurlumind\ as a luminosity
    indicator, and the SME results.
}
\end{deluxetable}

\subsection{Spectral line-bisector analysis}
\label{sec:bisec}

Our initial spectroscopic analyses discussed in \refsecl{recspec} and
\refsecl{hispec} rule out the most obvious astrophysical false positive
scenarios.  However, more subtle phenomena such as blends
(contamination by an unresolved eclipsing binary, whether in the
background or associated with the target) can still mimic both the
photometric and spectroscopic signatures we see.

Following \cite{torres:2007}, we explored the possibility that the
measured radial velocities are not real, but are instead caused by
distortions in the spectral line profiles due to contamination from a
nearby unresolved eclipsing binary \citep{queloz:2001}.  A bisector
analysis based on the Keck spectra was done as described in \S 5 of
\cite{bakos:2007a}.  We detect no correlated variation between the
bisector spans and the radial velocities (see \reffigl{rvbis}).  All
but two of the bisector measurements are consistent with no variation. 
Following the methods described in \cite{hartman:2009} and
\cite{kovacs:2010} we estimated the expected effect of contamination
from scattered moonlight on the bisectors, finding that the two outlier
measurements correspond to the two spectra that are expected to be the
most affected by sky contamination, when the Moon is $70^{\circ}$ from
the target.  Therefore, we conclude that the velocity variations are real, 
and that the star is orbited by a close-in giant planet.

\subsection{Global modeling of the data}
\label{sec:globmod}

This section briefly describes the procedure we followed to model the
HATNet photometry, the follow-up photometry, and the radial velocities
simultaneously.  More details on the fitting methods can be found in
\citet{bakos:2009}.  Our model for the follow-up \lcs\ used the
analytic formulae of \citet{mandel:2002} with quadratic limb darkening
coefficients interpolated from the tables by \citet{claret:2004}.
The transit shape was parametrized by the ratio-of-radii $p\equiv
\rpl/\rstar$, the square of the impact parameter $b^2$, and the
reciprocal of the half duration of the transit $\zrstar$.  We denote
this fitting set as \{$p$,$b^2$,$\zeta/R_*$\}.  This set is chosen
because of their simple geometric meanings and the fact that these 
exhibit low correlations \citep[see][]{bakos:2009}.
Our model for the HATNet data was the simplified ``P1P3'' version of
the \citet{mandel:2002} analytic functions following the method of
\citet{bakos:2009}.
%
Following the formalism presented by \citet{pal:2009}, the RVs were
fitted with an eccentric Keplerian model parametrized by the
semi-amplitude $K$ and Lagrangian elements $k \equiv e \cos\omega$ and
$h \equiv e \sin\omega$, in which $\omega$ is the longitude of
periastron.

Assuming a linear ephemeris, we assign the transit number $N_{tr} = 0$ 
to the complete follow-up \lc\ gathered on 2010 Feb 15.  
The eight main parameters describing the physical model were thus
$T_{c,-250}$, $T_{c,0}$, $\rpl/\rstar$, $b^2$, $\zrstar$, $K$, $k
\equiv e\cos\omega$, and $h \equiv e\sin\omega$.  Five additional
parameters were included that have to do with the instrumental
configuration.  These are the HATNet blend factors $B_{\rm inst,315}$
and $B_{\rm inst,316}$, which account for possible dilution of the
transit in the HATNet \lcs\ from background stars due to the broad PSF
(20\arcsec\ FWHM), the HATNet out-of-transit magnitudes $M_{\rm
0,HATNet,315}$ and $M_{\rm 0,HATNet,316}$, and the relative zero-point
$\gamma_{\rm rel}$ of the Keck RVs.

We extended our physical model with an instrumental model that
describes brightness variations caused by systematic errors in the
measurements as described in \citet{bakos:2009}.  The HATNet photometry
has already been EPD- and TFA-corrected before the global modeling, so
we only considered corrections for systematics in the follow-up \lcs. 
We chose the ``ELTG'' method, i.e., EPD was performed in ``local'' mode
with EPD coefficients defined for each night, and TFA was performed in
``global'' mode using the same set of stars and TFA coefficients for
all nights, as done in \citet{bakos:2009}.

The joint fit was accomplished using downhill simplex 
\citep[AMOEBA;see][]{press:1992} and the Markov Chain
Monte-Carlo method \citep[MCMC, see][]{ford:2006} using
``Hyperplane-CLLS'' chains \citep{bakos:2009} and the analytic partial
derivatives for the transit light curve from \citet{pal:2009}. 
A detailed description can be found in \citet{bakos:2009}.
The resulting geometric parameters pertaining to the light curves 
and velocity curves are listed in \reftabl{planetparam}. Quotes
values are the median and the error on the median from the
{\em a posteriori} distribution of each parameter.

Included in this table is the RV ``jitter'', which we added in
quadrature to the internal errors for the RVs in order to achieve
$\chi^{2}/{\rm dof} = 1$ from the RV data for the global fit. 
Auxiliary parameters not listed in the table are:
$T_{\mathrm{c},-250}=\hatcurLCTA$~(BJD),
$T_{\mathrm{c},0}=\hatcurLCTB$~(BJD), the blending factors 
$B_{\rm instr,314}=\hatcurLCiblendA$ and 
$B_{\rm instr,315}=\hatcurLCiblendB$, and 
$\gamma_{\rm rel}=\hatcurRVgammarel$\,\ms.
The latter quantity represents an arbitrary offset for the Keck RVs,
and does \emph{not} correspond to the true center of mass velocity of
the system, which was listed earlier in \reftabl{stellar} ($\gamma_{\rm
RV}$).

The planetary parameters and their uncertainties can be derived by
combining the {\em a posteriori} distributions for the stellar, \lc,
and RV parameters.  In this way we find a mass for the planet of
$\mpl=\hatcurPPmlong\,\mjup$ and a radius of
$\rpl=\hatcurPPrlong\,\rjup$, leading to a mean density
$\rho_p=\hatcurPPrho$\,\gcmc. These and other planetary parameters are
listed at the bottom of Table~\ref{tab:planetparam}.
We note that the system may be slightly eccentric: $e =
\hatcurRVeccen$, $\omega = \hatcurRVomega\arcdeg$.

\ifthenelse{\boolean{emulateapj}}{
    \begin{deluxetable*}{lr}
	\tabletypesize{\scriptsize}
}{
    \begin{deluxetable}{lr}
	\tabletypesize{\scriptsize}
}
\tablecaption{Orbital and planetary parameters\label{tab:planetparam}}
\tablehead{
	\colhead{~~~~~~~~~~~~~~~Parameter~~~~~~~~~~~~~~~} &
	\colhead{Value}
}
\startdata
\noalign{\vskip -5pt}
\sidehead{\Lc{} parameters}
~~~$P$ (days)             \dotfill    & $\hatcurLCP$              \\
~~~$T_c$ (${\rm BJD}$)    
      \tablenotemark{a}   \dotfill    & $\hatcurLCT$              \\
~~~$T_{14}$ (days)
      \tablenotemark{a}   \dotfill    & $\hatcurLCdur$            \\
~~~$T_{12} = T_{34}$ (days)
      \tablenotemark{a}   \dotfill    & $\hatcurLCingdur$         \\
~~~$\arstar$              \dotfill    & $\hatcurPPar$             \\
~~~$\zrstar$              \dotfill    & $\hatcurLCzeta$           \\
~~~$\rpl/\rstar$          \dotfill    & $\hatcurLCrprstar$        \\
~~~$b^2$                  \dotfill    & $\hatcurLCbsq$            \\
~~~$b \equiv a \cos i/\rstar$
                          \dotfill    & $\hatcurLCimp$            \\
~~~$i$ (deg)              \dotfill    & $\hatcurPPi$              \\
\noalign{\vskip -3pt}
\sidehead{Limb-darkening coefficients \tablenotemark{b}}
~~~$a_i$ (linear term)    \dotfill    & $\hatcurLBii$             \\
~~~$b_i$ (quadratic term) \dotfill    & $\hatcurLBiii$            \\
\noalign{\vskip -3pt}
\sidehead{RV parameters}
~~~$K$ (\ms)              \dotfill    & $\hatcurRVK$              \\
~~~$k_{\rm RV}$\tablenotemark{c} 
                          \dotfill    & $\hatcurRVk$              \\
~~~$h_{\rm RV}$\tablenotemark{c}
                          \dotfill    & $\hatcurRVh$              \\
~~~$e$                    \dotfill    & $\hatcurRVeccen$          \\
~~~$\omega$ (deg)         \dotfill    & $\hatcurRVomega$          \\
~~~RV jitter (\ms)        \dotfill    & 7.4           \\
\noalign{\vskip -3pt}
\sidehead{Secondary eclipse parameters}
~~~$T_s$ (BJD)            \dotfill    & $\hatcurXsecondary$       \\
~~~$T_{s,14}$             \dotfill    & $\hatcurXsecdur$          \\
~~~$T_{s,12}$             \dotfill    & $\hatcurXsecingdur$       \\
\noalign{\vskip -3pt}
\sidehead{Planetary parameters}
~~~$\mpl$ ($\mjup$)       \dotfill    & $\hatcurPPmlong$          \\
~~~$\rpl$ ($\rjup$)       \dotfill    & $\hatcurPPrlong$          \\
~~~$C(\mpl,\rpl)$
    \tablenotemark{d}     \dotfill    & $\hatcurPPmrcorr$         \\
~~~$\rhopl$ (\gcmc)       \dotfill    & $\hatcurPPrho$            \\
~~~$\log g_p$ (cgs)       \dotfill    & $\hatcurPPlogg$           \\
~~~$a$ (AU)               \dotfill    & $\hatcurPParel$           \\
~~~$T_{\rm eq}$ (K)       \dotfill    & $\hatcurPPteff$           \\
~~~$\Theta$\tablenotemark{e} \dotfill & $\hatcurPPtheta$          \\
~~~$F_{per}$ ($10^{\hatcurPPfluxperidim}$\ergscmsq)
                          \dotfill    & $\hatcurPPfluxperi$      \\
~~~$F_{ap}$  ($10^{\hatcurPPfluxapdim}$\ergscmsq) 
                          \dotfill    & $\hatcurPPfluxap$        \\
~~~$\langle F \rangle$ ($10^{\hatcurPPfluxavgdim}$\ergscmsq) \tablenotemark{f}
                          \dotfill    & $\hatcurPPfluxavg$        \\
[-1.5ex]
\enddata
\tablenotetext{a}{
    \ensuremath{T_c}: Reference epoch of mid transit that minimizes the
    correlation with the orbital period.  It corresponds to $N_{tr} =
    -8$. BJD is calculated from UTC.
	\ensuremath{T_{14}}: total transit duration.
	\ensuremath{T_{12}=T_{34}}: ingress/egress time.
}
\tablenotetext{b}{
	Values for a quadratic law, adopted from the tabulations by
    \cite{claret:2004} according to the spectroscopic (SME) parameters
    listed in \reftabl{stellar}. 
}
\tablenotetext{c}{
	\,Lagrangian orbital parameters derived from the global modeling, and
    primarily determined by the RV data.
}
\tablenotetext{d}{
	Correlation coefficient between the planetary mass \mpl\ and radius
	\rpl.
}
\tablenotetext{e}{
	The Safronov number is given by $\Theta = \frac{1}{2}(V_{\rm
	esc}/V_{\rm orb})^2 = (a/\rpl)(\mpl / \mstar )$
	\citep[see][]{hansen:2007}.
}
\tablenotetext{f}{
	Incoming flux per unit surface area, averaged over the orbit.
}
\ifthenelse{\boolean{emulateapj}}{
    \end{deluxetable*}
}{
    \end{deluxetable}
	\clearpage
}

\section{Comparison of Fitting Methods}
\label{sec:priors}

\subsection{Alternative fitting parameter sets}
\label{sec:fitsets}

The fitting method adopted for the values quoted in
Table~\ref{tab:planetparam} uses the parameter set
\{$p$,$\zeta/R_*$,$b^2$\}, chosen for their low inter-parameter
correlations.  $\zeta/R_*$ is the reciprocal of the half-duration as
computed using an approximate expression for the duration coming from
\citet{tingley:2005}.  Recently, \citet{kipping:2010} showed that an
improved approximate formula is possible for the duration.  Replacing
$\zeta/R_*$ with the reciprocal of the new expression for the
half-duration therefore offers greater accuracy for the duration
determination.  \citet{kipping:2010} labeled this parameter as
$\Upsilon/R_*$ to be distinct from $\zeta/R_*$.  The new parameter set
yields the greatest improvements for near-grazing, low-eccentricity
orbits.

To investigate the robustness of the results against different
parameter sets, we refitted the EPD, TFA corrected HAT and FLWO light
curves in conjunction with the radial velocities using the
\{$p^2$,$b^2$,$\Upsilon/R_*$\} parameter set.  Radial velocities are
fitted using the Lagrange parameters as in \refsecl{globmod}.  A new YY
\citep{yi:2001} isochrone analysis is performed as described in
\refsecl{stelparam} to show the effect on the physical parameters.  
The results are shown in the first results column of
Table~\ref{tab:global}.

The $b^2$ parameter also gives us some pause for thought. One natural
reason to select $b^2$ is that the duration of a transit is completely
described in terms of $b^2$ i.e.~there is no case of an isolated $b$
which is not squared in the expression for the duration.  Therefore,
$b^2$ seems to be a natural parameter of the transit light curve. 
Further, $b^2$ usually shows lower inter-parameter correlations than
$b$.

However, there are also two reasons why one should not choose $b^2$.
Firstly, geometrically $b$ is more likely to have a uniform prior than
$b^2$.  Therefore, low signal-to-noise transits will be biased towards
higher impact parameters by fitting for $b^2$.  Secondly, $b^2$ cannot
be negative and thus the posterior distribution of $b^2$ tends to get
offset to an artificially more positive value due to the boundary
condition that $b^2 > 0$.  One resolution to this is to use $b$ and let
the parameter explore both negative and positive values.  Whilst a
negative impact parameter may seem unphysical, the issue of its
physicality is also irrelevant since the transit is completely
described in terms of $b^2$ and thus a negative $b$ is always
multiplied by itself when computing the light curve morphology.  Using
$b$ in this way permits for a symmetric distribution about $b=0$ and
thus improved estimates of the associated uncertainty.

To investigate the value of fitting for $b$, we repeated our global
fits using the parameter set \{$p^2$,$b$,$T_{1.5,3.5}$\}. 
$T_{1.5,3.5}$ replaces $\Upsilon/R_*$ since although it may be slightly
more correlated, it is more reasonable to expect a uniform prior on the
duration than a uniform prior on its reciprocal.  The results of these
fits are shown in the third results column of Table~\ref{tab:global}.

The system values quoted between \{$p^2$,$b^2,\zeta/R_*$\},
\{$p^2$,$b^2,\Upsilon/R_*$\} and \{$p^2$,$b$,$T_{1.5,3.5}$\} show
excellent agreement and thus indicate that the system parameters are
insensitive to the choice of priors.

\subsection{Fitted limb darkening}
\label{sec:ldfit}

We repeated the fits for the \{$p^2$,$b^2,\Upsilon/R_*$\} and
\{$p^2$,$b$,$T_{1.5,3.5}$\} parameter sets with fitted limb darkening
(second and fourth result columns of Table~\ref{tab:global}
respectively).  In order to achieve convergence, we choose to only use
linear limb darkening and thus fix the quadratic coefficient to be
zero.

The results show slight differences with the fixed limb darkening
analogues.  The most noticeable effect is increased error bars. 
Fitting for limb darkening means that our results are no longer
dependent upon a stellar atmosphere model prediction, which is a clear
desideratum.  Our preferred final values are given by the
\{$p^2$,$b$,$T_{1.5,3.5}$\} parameter set with fitted limb darkening
(last column of Table~\ref{tab:global}, with highlighted header).

Both free limb darkening fits converge at $u_1 = 0.25 \pm 0.04$. For
$0.8<\mu<1$, the limb darkening from both the theoretical quadratic
coefficients and the fitted coefficients is approximately equivalent. 
However, for low $\mu$, the theoretical coefficients predict much
stronger darkening effects than observed.  Fitting a linear law through
the quadratic coefficients gives $u_{\mathrm{lin}} = 0.46$,
demonstrating the stronger limb darkening predicted from theory.


\ifthenelse{\boolean{emulateapj}}{
    \begin{deluxetable*}{lrrrr}
}{
    \begin{deluxetable}{lrrrr}
}
\tablewidth{0pc}
\tabletypesize{\scriptsize}
\tablecaption{Orbital and planetary parameters\label{tab:global}}
\tablehead{
	\colhead{~~~~~~~~~~~~~~~Parameter~~~~~~~~~~~~~~~} &
	\colhead{\{$b^2$,$\Upsilon/R_*$\}, fixed LD} &
	\colhead{\{$b^2$,$\Upsilon/R_*$\}, fitted LD} &
	\colhead{\{$b$,$T_{1.5,3.5}$\}, fixed LD} &
	\colhead{\{$\mathbf{b}$,$\mathbf{T_{1.5,3.5}}$\}, \textbf{fitted LD} }
}
\startdata
\noalign{\vskip -3pt}
\sidehead{\Lc{} parameters}
$P$ (days) \dotfill & $3.3552459_{-0.0000072}^{+0.0000079}$ & $3.3552466_{-0.0000070}^{+0.0000069}$ & $3.3552458_{-0.0000072}^{+0.0000078}$ & $\mathbf{3.3552464_{-0.0000071}^{+0.0000069}}$ \\
$T_c$ (BJD$_{\mathrm{UTC}}$ - 2,450,000) \tablenotemark{a} \dotfill & $5216.97672_{-0.00025}^{+0.00025}$ & $5216.97668_{-0.00024}^{+0.00024}$ & $5216.97672_{-0.00025}^{+0.00024}$ & $\mathbf{5216.97669_{-0.00024}^{+0.00024}}$ \\
$T_{1,4}$ (s) \tablenotemark{a} \dotfill & $13307.3_{-55.2}^{+62.0}$ & $13172.1_{-79.3}^{+100.5}$ & $13307.8_{-53.8}^{+58.7}$ & $\mathbf{13152.7_{-69.2}^{+88.6}}$ \\
$T_{1.5,3.5}$ (s) \tablenotemark{a} \dotfill & $12112.1_{-44.5}^{+44.0}$ & $11908.8_{-52.6}^{+53.4}$ & $12111.8_{-44.6}^{+45.1}$ & $\mathbf{11912.1_{-52.7}^{+52.7}}$ \\
$T_{2,3}$ (s) \tablenotemark{a} \dotfill & $10917.0_{-53.9}^{+46.3}$ & $10646.4_{-139.6}^{+97.2}$ & $10915.8_{-49.0}^{+45.1}$ & $\mathbf{10676.6_{-121.0}^{+75.7}}$ \\
$T_{1,2} \simeq T_{3,4}$ (s) \tablenotemark{a} \dotfill & $1184.4_{-11.3}^{+47.2}$ & $1259.5_{-75.6}^{+110.8}$ & $1189.3_{-12.9}^{+32.3}$ & $\mathbf{1227.4_{-41.9}^{+101.3}}$ \\
$(R_P/R_*)^2$ (\%) \dotfill & $0.9363_{-0.0092}^{+0.0093}$ & $0.978_{-0.014}^{+0.015}$ & $0.9362_{-0.0090}^{+0.0091}$ & $\mathbf{0.976_{-0.013}^{+0.014}}$ \\
$R_P/R_*$ \dotfill & $0.09676_{-0.00047}^{+0.00048}$ & $ 0.09889_{-0.00071}^{+0.00073}$ & $0.09676_{-0.00047}^{+0.00047}$ & $\mathbf{0.09877_{-0.00068}^{+0.00069}}$ \\
$a/R_*$ \dotfill & $7.71_{-0.31}^{+0.32}$ & $7.63_{-0.38}^{+0.37}$ & $7.71_{-0.30}^{+0.31}$ & $\mathbf{7.70_{-0.36}^{+0.35}}$ \\
$\Upsilon/R_*$ (days$^{-1}$) \dotfill & $14.267_{-0.052}^{+0.053}$ & $14.510_{-0.065}^{+0.064}$ & $14.267_{-0.053}^{+0.053}$ & $\mathbf{14.506_{-0.064}^{+0.064}}$ \\
$b$ \dotfill & $0.00_{-0.00}^{+0.20}$ & $0.25_{-0.24}^{+0.12}$ & $0.00_{-0.13}^{+0.13}$ & $\mathbf{0.00_{-0.26}^{+0.26}}$ \\
$b^2$ \dotfill & $0.000_{-0.000}^{+0.041}$ & $0.060_{-0.060}^{+0.072}$ & $0.008_{-0.007}^{+0.025}$ & $\mathbf{0.037_{-0.033}^{+0.070}}$ \\
$i$ (deg) \dotfill & $90.0_{-1.5}^{+0.0}$ & $88.2_{-1.0}^{+1.8}$ & $90.0_{-1.0}^{+1.0}$ & $\mathbf{90.0_{-1.9}^{+1.9}}$ \\
$\rho_*$ (g\,cm$^{-3}$) \dotfill & $0.770_{-0.088}^{+0.099}$ & $0.748_{-0.105}^{+0.114}$ & $0.771_{-0.086}^{+0.097}$ & $\mathbf{0.768_{-0.102}^{+0.110}}$ \\
\sidehead{Limb-darkening coefficients \tablenotemark{b}}
$u_1$ (linear term)  \dotfill & $0.1858^{*}$ & $0.249_{-0.039}^{+0.038}$ & $0.1858^{*}$ & $\mathbf{0.251_{-0.038}^{+0.037}}$ \\
$u_2$ (quadratic term)  \dotfill & $0.3625^{*}$ & $0^{*}$ & $0.3625^{*}$ & $\mathbf{0^{*}}$ \\
\sidehead{RV derived parameters}
$k_{\rm RV}$\tablenotemark{c} \dotfill & $-0.037_{-0.014}^{+0.014}$ & $-0.037_{-0.014}^{+0.014}$ & $-0.037_{-0.014}^{+0.014}$ & $\mathbf{-0.037_{-0.014}^{+0.014}}$ \\
$h_{\rm RV}$\tablenotemark{c} \dotfill & $-0.018_{-0.039}^{+0.038}$ & $-0.017_{-0.039}^{+0.038}$ & $-0.018_{-0.038}^{+0.038}$ & $\mathbf{-0.018_{-0.039}^{+0.038}}$ \\
$\Psi$\tablenotemark{c} \dotfill & $1.007_{-0.004}^{+0.012}$ & $1.006_{-0.004}^{+0.012}$ & $1.006_{-0.004}^{+0.012}$ & $\mathbf{1.007_{-0.004}^{+0.012}}$ \\
$K$ (\ms) \dotfill & $82.6_{-3.2}^{+3.1}$ & $82.6_{-3.2}^{+3.1}$ & $82.6_{-3.1}^{+3.1}$ & $\mathbf{82.6_{-3.1}^{+3.2}}$ \\
$e$ \dotfill & $0.052_{-0.017}^{+0.022}$ & $0.052_{-0.017}^{+0.022}$ & $0.052_{-0.017}^{+0.022}$ & $\mathbf{0.052_{-0.017}^{+0.022}}$ \\
$\omega$ (deg) \dotfill & $206_{-54}^{+35}$ & $205_{-54}^{+36}$ & $206_{-54}^{+35}$ & $\mathbf{206_{-53}^{+36}}$ \\
RV jitter (\ms) \dotfill & $7.53_{-7.53}^{+3.60}$ & $7.44_{-7.44}^{+3.54}$ & $7.58_{-7.58}^{+3.65}$ & $\mathbf{7.43_{-7.43}^{+3.53}}$ \\
$\log g_p$ (cgs) \dotfill & $3.055_{-0.040}^{+0.039}$ & $3.027_{-0.050}^{+0.046}$ & $3.055_{-0.038}^{+0.039}$ & $\mathbf{3.037_{-0.047}^{+0.043}}$ \\
\sidehead{Secondary eclipse parameters}
$T_{S}$ (BJD$_{\mathrm{UTC}}$ - 2,450,000) & $5218.575_{-0.030}^{+0.030}$ & $55218.575_{-0.030}^{+0.030}$ & $5218.576_{-0.030}^{+0.029}$ & $\mathbf{5218.575_{-0.030}^{+0.030}}$ \\
$T_{S,1,4}$ (s) & $12850_{-940}^{+1000}$ & $12770 _{-910}^{+930}$ & $12850_{-940}^{+990}$ & $\mathbf{12730_{-920}^{+950}}$ \\
\sidehead{Stellar parameters}      
$M_*$ ($M_{\odot}$) \dotfill & $1.184_{-0.039}^{+0.041}$ & $1.189_{-0.041}^{+0.041}$ & $1.184_{-0.039}^{+0.040}$ & $\mathbf{1.186_{-0.041}^{+0.042}}$ \\ 
$R_*$ ($R_{\odot}$) \dotfill & $1.293_{-0.057}^{+0.061}$ & $1.307_{-0.066}^{+0.077}$ & $1.293_{-0.056}^{+0.058}$ & $\mathbf{1.294_{-0.062}^{+0.071}}$ \\
log$(g_*)$ (cgs) \dotfill & $4.287_{-0.034}^{+0.034}$ & $4.279_{-0.042}^{+0.040}$ & $4.287_{-0.033}^{+0.034}$ & $\mathbf{4.286_{-0.039}^{+0.038}}$ \\ 
$L_*$ ($L_{\odot}$) \dotfill & $2.47_{-0.26}^{+0.29}$ & $2.52_{-0.29}^{+0.35}$ & $2.47_{-0.25}^{+0.28}$ & $\mathbf{2.48_{-0.28}^{+0.32}}$ \\
$M_{V}$ (mag) \dotfill & $3.78_{-0.13}^{+0.13}$ & $3.76_{-0.14}^{+0.14}$ & $3.78_{-0.12}^{+0.13}$ & $\mathbf{3.78_{-0.14}^{+0.14}}$ \\
Age (Gyr) \dotfill & $2.75_{-0.64}^{+0.58}$ & $2.80_{-0.63}^{+0.55}$ & $2.75_{-0.64}^{+0.58}$ & $\mathbf{2.75_{-0.65}^{+0.58}}$ \\
Distance (pc) \dotfill & $404_{-23}^{+24}$ & $409_{-26}^{+28}$ & $404_{-23}^{+24}$ & $\mathbf{405_{-25}^{+27}}$ \\
\sidehead{Planetary parameters}
$M_P$ ($M_J$) \dotfill & $0.680_{-0.030}^{+0.030}$ & $0.682_{-0.031}^{+0.031}$ & $0.680_{-0.030}^{+0.030}$ & $\mathbf{0.681_{-0.030}^{+0.031}}$ \\ 
$R_P$ ($R_J$) \dotfill & $1.217_{-0.054}^{+0.0581}$ & $1.257_{-0.066}^{+0.079}$ & $1.217_{-0.053}^{+0.056}$ & $\mathbf{1.243_{-0.061}^{+0.072}}$ \\
C\{$M_P$/$R_P$\} \tablenotemark{d} \dotfill & $0.264$ & $0.300$ & $0.266$ & $\mathbf{0.290}$ \\
$\rho_P$ (g\,cm$^{-3}$) \dotfill & $0.467_{-0.059}^{+0.066}$ & $0.425_{-0.068}^{+0.073}$ & $0.468_{-0.057}^{+0.065}$ & $\mathbf{0.439_{-0.066}^{+0.069}}$ \\ 
$a$ (AU) \dotfill & $0.04639_{-0.00052}^{+0.00052}$ & $0.04645_{-0.00055}^{+0.00053}$ & $0.04638_{-0.00052}^{+0.00052}$ & $\mathbf{0.04641_{-0.00054}^{+0.00054}}$ \\
$T_{\mathrm{eq}}$ (K) \dotfill & $1623_{-38}^{+39}$ & $1632_{-43}^{+47}$ & $1623_{-37}^{+38}$ & $\mathbf{1624_{-41}^{+44}}$ \\
$\Theta$ \tablenotemark{e} \dotfill & $0.0761_{-0.0031}^{+0.0032}$ & $0.0737_{-0.0039}^{+0.0038}$ & $0.0761_{-0.0030}^{+0.0031}$ & $\mathbf{0.0745_{-0.0037}^{+0.0035}}$ \\
$F_{\mathrm{per}}$ ($10^{\hatcurPPfluxperidim}$\ergscmsq) \tablenotemark{f} \dotfill & $1.73_{-0.12}^{+0.17}$ & $1.77_{-0.16}^{+0.21}$ & $1.73_{-0.12}^{+0.17}$ & $\mathbf{1.74_{-0.14}^{+0.20}}$ \\
$F_{\mathrm{ap}}$ ($10^{\hatcurPPfluxapdim}$\ergscmsq) \tablenotemark{f} \dotfill & $1.43_{-0.17}^{+0.14}$ & $1.46_{-0.18}^{+0.18}$ & $1.43_{-0.17}^{+0.14}$ & $\mathbf{1.43_{-0.18}^{+0.16}}$ \\
$\langle F \rangle$ ($10^{\hatcurPPfluxavgdim}$\ergscmsq) \tablenotemark{f} \dotfill & $1.57_{-0.14}^{+0.16}$ & $1.60_{-0.16}^{+0.19}$ & $1.57_{-0.14}^{+0.15}$ & $\mathbf{1.57_{-0.15}^{+0.18}}$ \\
[-0.5ex]
\enddata
\tablenotetext{a}{
    \ensuremath{T_c}: Reference epoch of mid transit that minimizes the
    correlation with the orbital period.  It corresponds to $N_{tr} =
    -8$.
	\ensuremath{T_{x,y}}: transit duration between contact points $x$
	and $y$. $(x,y)=(1.5,3.5)$ correspond to the sky-projected
        center of the planet overlapping the stellar limb.
}
\tablenotetext{b}{
	Values for a quadratic law and fixed coefficients, adopted from the
    tabulations by \cite{claret:2004} according to the spectroscopic
    (SME) parameters listed in \reftabl{stellar}.
}
\tablenotetext{c}{
    \ensuremath{k \& h}: Lagrangian orbital parameters derived from the
    global modeling, and primarily determined by the RV data.
	\ensuremath{\Psi}: Reciprocal of the scaling factor by which the
        true stellar density is modified from that found assuming a
        circular orbit \citep{kipping:2010}.
}
\tablenotetext{d}{
	Correlation coefficient between the planetary mass \mpl\ and radius
	\rpl.
}
\tablenotetext{e}{
	The Safronov number is given by $\Theta = \frac{1}{2}(V_{\rm
	esc}/V_{\rm orb})^2 = (a/\rpl)(\mpl / \mstar )$
	\citep[see][]{hansen:2007}.
}
\tablenotetext{f}{
	Incoming flux per unit surface area, averaged over the orbit.
}
\ifthenelse{\boolean{emulateapj}}{
    \end{deluxetable*}
}{
    \end{deluxetable}
	\clearpage
}

\section{Orbital Eccentricity}
\label{sec:ecc}
\subsection{Significance of the eccentric fit}
\label{sec:eccsig}

The eccentric fit suggests an orbital eccentricity of $e =
0.052_{-0.017}^{+0.022}$.  In addition to the four fits displayed in
Table~\ref{tab:global}, we repeated our preferred model fit (i.e.~
\{$p^2$,$b$,$T_{1.5,3.5}$\} with fitted limb darkening) for a circular
orbit.  This was done to provide a $\chi^2$ value for both fits, which
can be used to infer the statistical significance of the eccentric 
fit.

To accomplish this, we only take the $\chi^2$ from the radial velocity
data, which dominates the determination of the Lagrangian orbital
parameters and so the number of data points is $n=18$.  The period and
time of transit are dominated by the photometry and very weakly
affected by the few RV points.  Therefore, these two degrees of 
freedom can be considered fixed.  This leaves us with four degrees of 
freedom for an eccentric fit ($k$, $h$, $K$, $\gamma$) and two for the 
circular fit ($K$, $\gamma$).

In evaluating the significance of the eccentric fit over the circular
model it is important to penalize the eccentric fit for using two 
extra degrees of freedom.  We therefore choose to perform an F-test 
between the two models.  The circular orbit fit has $\chi^2 = 59.5$ 
and the eccentric fit has $\chi^2 = 39.6$.  The false alarm 
probability from an F-test is evaluated to be 5.8\% or 1.9-$\sigma$.  
We also performed the test of \citet{lucy71}, where the statistical 
significance of the eccentric fit is given by:

\begin{equation}
\mathrm{P}(e > 0) = 1 - \exp\Big[-\frac{\hat{e}^2}{2 \sigma_e^2}\Big]
\end{equation}

Where $\hat{e}$ is the modal value of the eccentricity, which is well
approximated by the median for a unimode distribution.  Using the
\citet{lucy71} test, we find an eccentric fit is accepted at the
2.6-$\sigma$ level.  The slightly higher significance likely comes 
from the fact this test does not penalize an eccentric model for using 
more degrees of freedom, whereas the F-test does.  Therefore, based 
upon the current data for this system, an eccentric fit is probable 
but not conclusive.

A resolution would be to obtain a secondary eclipse measurement for the
system, for which the mid-eclipse time would be dependent upon
$e\cos\omega$.  The $k$ component dominates the eccentricity budget and
thus its determination would strongly constrain the eccentricity of
this system.

\subsection{Circularization timescales}
\label{sec:circularization}

Tidal dissipation causes planetary orbits to circularize over time. The
maximum eccentricity a planet could initially have is $e \sim 1$. 
After $N=1$ circularization timescales, denoted $\tau_{\mathrm{circ}}$,
the planet's eccentricity will reduce by one e-fold, i.e.~a factor of
2.72.  For the planet to now have an eccentricity of $e$, the number of
circularization timescales which have transpired must be $\leq
-\log(e)$ and therefore $T_{\mathrm{Age}} \leq -\log(e)
\tau_{\mathrm{circ}}$.  This therefore constrains the circularization
timescale to be:

\begin{equation}
\tau_{\mathrm{circ}} \geq \frac{T_{\mathrm{Age}}}{-\log(e)}
\end{equation}

Using the method of \citet{ada06}, the circularization timescale may be
expressed as a function of the planet's tidal dissipative constant,
$Q_P$, for a low eccentricity system.

\begin{equation}
\tau_{\mathrm{circ}} = Q_P \frac{4}{63} \frac{P}{2\pi} \frac{M_P}{M_*} \Big(\frac{a}{R_*} \frac{1}{p}\Big)^5 (1-e^2)^{13/2}
\end{equation}

Taking $e \simeq 0$ in the above expression together with the posterior
distributions of the various parameters given above allows us to
constrain $Q_P$ to be $Q_P \geq (6.1_{-2.0}^{+3.0}) \times 10^6$.  
Note that the value above becomes even larger if we include the
$(1-e^2)^{13/2}$ term from the \citet{ada06} equation, but this
requires some assumption of the history of the system.  Given than
Jupiter has $Q_P \sim 30,000$ \citep{lainey:2009}, this limit raises
some questions about why HAT-P-24b has such a large value, somewhat
similar to the situation for GJ 436b \citep{deming:2007}.  In none of
our $10^5$ realizations do we have a $Q_P$ value below 150,000 and
therefore in the absence of any eccentricity pumping, a large $Q_P$
value is a possible origin for the non-zero eccentricity.  This is
consistent with the observation of large $Q_P$ values in many other
known TEPs \citep{matsumura:2008}.

We note that for planets with initial eccentricities $\geq 0.2$,
the above approximate expressions will be invalid and a full backwards
integration of the planet's orbital evolution will be necessary, as
pointed out by \citet{leconte:2010}.  Such a detailed analysis remains
outside of the scope of this paper, but our calculations do flag this
system as possibly retaining an anomalously large eccentricity 
requiring further investigation.

\section{Linear RV Drift}
\label{sec:drift}

The unfolded residuals of an eccentric fit seem to hint at a negative
linear drift in the radial velocities.  We re-executed the global fit
of the data including a RV gradient term $\dot{\gamma}$.  We choose to
use the \{$p^2$,$b$,$T_{1.5,3.5}$\} parameter set with fitted limb
darkening again.  The fits obtain $\dot{\gamma} =
-0.040_{-0.028}^{+0.028}$\,m/s/day with a slightly decreased
eccentricity of $e = 0.048_{-0.017}^{+0.022}$.  By the \citet{lucy71}
test, the eccentricity is now significant at the 2.3-$\sigma$
confidence level.  We also note that the $\dot{\gamma}$ parameter
appears to have converged in the MCMC trials with the Gelman-Rubins
statistic \citep{gel92} satisfying the criteria of being $<1$ (value
was 0.59), indicative of good-mixing.

\subsection{Statistical significance}
\label{sec:driftsig}

\subsubsection{F-test}

There are numerous tests which one can employ to evaluate the
significance of the gradient.  The first one we tried was to compute
the F-test between the eccentric orbit and the eccentric orbit + linear
drift model.  Penalizing for one extra degree of freedom, the F-test
find the drift model is accepted with $76.4$\% confidence.

\subsubsection{Odds ratio}

The second test we tried was to extract the posterior distribution of
the gradient from the MCMC runs.  If the gradient was equal to zero, we
would expect 50\% of the MCMC runs to give a positive value and 50\% to
give a negative value.  In the eccentric + drift model,
$f_{\mathrm{neg}} = 92.7$\% of the MCMC runs gave a negative
$\dot{\gamma}$.  The odds ratio of the negative valued model over the
50:50 model is:

\begin{equation}
\mathrm{O}_{\mathrm{drift/static}} = \frac{0.5}{1-f_{\mathrm{neg}}}
\end{equation}

For only two possible models (i.e.~a drift or static), the probability
of the drift model being the correct one is $\mathrm{P}(\mathrm{drift})
= 1 - [1/(1 + \mathrm{O}_{\mathrm{drift/static}})] = 87.2$\%. 
Therefore, both tests so far indicate a $\sim20$\% false alarm
probability for the drift model.

\subsubsection{Bayesian Information Criterion}
\label{sec:bic}

The final test we performed was to re-fit all of the data using four
possible models, using the \{$p^2$,$b$,$T_{1.5,3.5}$\} and fitted limb
darkening method, each with a different number of degrees of freedom,
$d$:

\begin{enumerate}
\item Circular orbit ($d=2$)
\item Circular orbit + linear drift ($d=3$)
\item Eccentric orbit ($d=4$)
\item Eccentric orbit + linear drift ($d=5$)
\end{enumerate}

In each case we compute the Bayesian Information Criterion (BIC)
(\citet{schwarz:1978}; \citet{liddle:2007}), given by BIC$= \chi^2 + d
\log n$ where $n$ is the number of RV data points.  BIC severely
penalizes models for having more parameters and offers a statistically
valid tool for model selection.  We also compute the reduced $\chi^2$,
given by $\chi_{\mathrm{reduced}}^2 = \chi^2/(n-d)$.  As an example,
the circular orbit has only two degrees of freedom in $\gamma$ and $K$. 
One might argue that $P$ and $t_C$ are also degrees of freedom but in
a global fit, which includes the HAT and FLWO time series,
these two parameters are overwhelmingly driven by the photometry and
not the RV and thus the RV actually has negligible freedom in these
parameters.

\ifthenelse{\boolean{emulateapj}}{
    \begin{deluxetable*}{lccccc}
}{
    \begin{deluxetable}{lccccc}
}
\tablewidth{0pc}
\tabletypesize{\scriptsize}
\tablecaption{
	Comparison of four different models for the RVs of HAT-P-24b, described
	in \refsecl{bic}. 
	\label{tab:lindrift}
}
\tablehead{
	\colhead{Model} & 
	\colhead{$d$} & 
	\colhead{$\chi_{\mathrm{reduced}}^2$} &
	\colhead{BIC} & 
	\colhead{$\dot{\gamma}$} & 
	\colhead{$e$} \\
	\colhead{} & 
	\colhead{} & 
	\colhead{} &
	\colhead{} & 
	\colhead{$\mathrm{m\,s^{-1}day^{-1}}$} & 
	\colhead{}
}
\startdata
Circular \dotfill & 2 & 3.72 & 65.3 & $0$\tablenotemark{a} & $0$\tablenotemark{a} \\
Circular + Drift \dotfill & 3 & 3.35 & 59.0 & $-0.053_{-0.026}^{+0.026}$ & $0$\tablenotemark{a} \\
Eccentric \dotfill & 4 & 2.83 & 51.2 & $0$\tablenotemark{a} & $0.052_{-0.017}^{+0.022}$ \\
Eccentric + Drift \dotfill & 5 & 2.72 & 49.9 & $-0.040_{-0.028}^{+0.028}$ & $0.048_{-0.017}^{+0.022}$ \\
\enddata
\tablenotetext{a}{
	Parameter is fixed.
}
\ifthenelse{\boolean{emulateapj}}{
    \end{deluxetable*}
}{
    \end{deluxetable}
}

The BIC model selection test indicates that the eccentric orbit +
linear drift model is the accepted model description of the current
radial velocities when globally fitted with the current photometry for
this system.  We note that i) the eccentric models are consistently
preferred over the circular orbit models ii) the eccentricity is
affected by a negligible degree by including the drift.

\subsubsection{Conclusion}

The linear drift model is the preferred model using the Bayesian
Information Criterion.  Other tests indicate the model is accepted with
a false alarm probability of 20\%.  This is not sufficient to yet claim
the trend is real and thus we encourage observers to obtain more
observations to confirm or reject the existence of this trend.

\subsection{Properties of Putative HAT-P-24c}
\label{sec:hatp24c}

We proceed here to constrain the properties of HAT-P-24c under the
assumption the trend is real.  Whilst this may not turn out to be true,
it is useful to consider what the properties of the outer planet would
be should the trend be later confirmed.

The period of the outer planet would have to be much greater than the
timescale of the observations, or a sinusoidal pattern would have
emerged and so $P_c \gg 265$\,days, most likely of
$\mathcal{O}\sim1000$\,days which constrains $a_c \gtrsim 2$\,AU by
Kepler's Third Law.  We note that the habitable zone pushes out as
$a_{\mathrm{hab}} = \sqrt{L_*/L_{\odot}}$\,AU and occurs at
$\sim1.6$\,AU for HAT-P-24 and so the outer planet would likely be
``cold''.  Using equation (1) from \citet{winn:2009}, the gradient
corresponds to an outer planet satisfying:

\begin{equation}
\frac{M_c \sin i_c}{a_c^2} = \frac{|\dot{\gamma}|}{G} = 0.082_{-0.056}^{+0.051} \,M_{J} \mathrm{AU}^{-2}
\end{equation}

Based upon the $a_c$ constraint, this therefore implies $M_c \sin i_c
\gtrsim 0.3$\,$M_J$.  Aside from the RVs, there are observational
consequences for HAT-P-24b due to the outer planet.  The system will
behave as an inner-outer binary and thus the outer planet will induce a
light-time travel effect, potentially detectable with transit timing
variations (TTV) of HAT-P-24b.  The inner binary will orbit the
barycentre with semi-major axis $(a_c M_c \sin i_c)/(M_*+M_b)$. 
Therefore the peak-to-peak light-time effect, for an outer planet of
negligible eccentricity, will be:

\begin{equation}
\mathrm{TTV}_{\mathrm{light}}(\mathrm{peak-to-peak}) \simeq \frac{|\dot{\gamma}| }{c} \frac{P_c^2}{2\pi^2}
\end{equation}

Adding in the best-fit value of $\dot{\gamma}$ gives
$\mathrm{TTV}_{\mathrm{light}}(\mathrm{peak-to-peak}) = 0.078
(P_c/\mathrm{years})^2$\,s.  Therefore, we require $P_c > 3.6$\,years
for a $>1$\,second TTV and would need $P_c > 36$\,years for a $>100$\,s
TTV.  We also evaluated the TTV effect due to a distant, perturbing
planet, as described in case IV of \citet{agol:2005}.  We find that a
1000\,day period 0.3\,$M_J$ would generate an r.m.s.~TTV of 0.008\,s
for $e_c = 0.1$, 0.06\,s for $e_c=0.5$ and 0.8\,s for $e_c=0.9$.  The
challenge of measuring TTVs to this precision over such long
time-scales is a daunting one and unlikely to reap any reward with
current instrumentation.

\section{Transit times}
\label{sec:tmids}

Whilst the photometric quality of the FLWO data is sufficient for full
free fitting, three of the four FLWO light curves are only partial
transits and so the errors on the duration and therefore mid-transit
time diverge for unconstrained fitting parameters.

A solution to this is to work under the assumption that the duration
and depth of the transit do not change from transit-to-transit.  As a
result of the partial transits, a transit duration variation (TDV)
analysis is therefore not possible but TTVs can be obtained provided it
is understood that they are derived as inherently model dependent
values, where the model is that of constant duration and depth.

We therefore extract the parameters $p^2$, $b$, $T$, $e\sin\omega$,
$e\cos\omega$, $P$ and the linear limb darkening coefficient $u_1$ from
the posterior distribution of the global fit.  We select the
\{$b$,$T$\} parameter set with fitted limb darkening as our favorite
solution for this purpose.  The free parameters of the individual
transit fits are OOT and $t_C$.  We stress that the parameters assumed
to be constant from transit-to-transit are still allowed to float
around their median-value with standard deviation given by their
derived uncertainties.  This ensures the errors are correctly
propagated into the mid-times.  The final times are given in
Table~\ref{tab:tmids}.

Using the linear ephemeris derived from the global fit, including all
HAT data, we find that the FLWO transits show no excess variance
yielding $\chi^2 = 1.3$ for 2 degrees of freedom.  The RMS of the
four O-C values is 34.7\,seconds.  With only four transit times, it is
not possible to conduct a meaningful TTV analysis.  However, these
transits may be used a benchmark for future TTV searches on this
system.

\ifthenelse{\boolean{emulateapj}}{
    \begin{deluxetable}{lrr}
}{
    \begin{deluxetable}{lrr}
}
\tablewidth{0pc}
\tabletypesize{\scriptsize}
\tablecaption{
	Fitted mid-transit times.
	\label{tab:tmids}
}
\tablehead{
	\colhead{Epoch} & 
	\colhead{$t_C$/(BJD$_{\mathrm{UTC}}$-2,450,000)} & 
	\colhead{O-C}\\
	\colhead{} & 
	\colhead{days} & 
	\colhead{seconds}
}
\startdata
-11 & $5206.91085_{-0.00062}^{+0.00062}$ & $-6.9_{-53.6}^{+53.6}$ \\
-9 & $5213.62078_{-0.00065}^{+0.00065}$ & $-55.6_{-56.2}^{+56.2}$ \\
-3 & $5233.75317_{-0.00082}^{+0.00080}$ & $23.1_{-70.8}^{+69.1}$ \\
0 & $5243.81878_{-0.00030}^{+0.00030}$ & $11.9_{-26.0}^{+26.0}$ \\ 
\enddata
\ifthenelse{\boolean{emulateapj}}{
    \end{deluxetable}
}{
    \end{deluxetable}
}

\section{Follow-up Possibilities}
\label{sec:followup}

\subsection{Secondary Eclipse}
\label{sec:sececlipse}

The Keck RVs indicate that $e\cos\omega$ is significantly non-zero and
thus suggests the secondary eclipse of HAT-P-24b would occur with a
timing offset from that a circular orbit.  Neglecting terms of order
$\cot^2i$, \citet{sterne:1940} shows that the timing offset is given
by:

\begin{equation}
\Delta t = \frac{P}{\pi} \Bigg(\frac{e\cos\omega \sqrt{1-e^2}}{(1-(e\sin\omega)^2)} + \arctan\Big(\frac{e\cos\omega}{\sqrt{1-e^2}}\Big)\Bigg)
\end{equation}

Using the posterior distribution of parameters from the
\{$p^2$,$b$,$T_{1.5,3.5}$\} fitted-limb-darkening MCMC run, we estimate
that the secondary eclipse should occur $(1.9 \pm 0.7)$\,hours earlier
than that expected for a purely circular orbit.  The detection of the
eclipse would therefore strongly constrain $e\cos\omega$ which would
allow for a revised global fit to the data.  Constraining $e\cos\omega$
in this way would allow us to re-compute the statistical significance
of the RV linear drift model over the static model.

For \emph{Spitzer}'s 3.6\,$\mu$m and 4.5\,$\mu$m channels, assuming
uniform redistribution of energy around the planet and zero albedo, we
estimate a depth of $\sim 0.075$\% and $\sim0.10$\%.  At $K_s =
10.5$\,mag, the brightness is comparable to that of TrES-4 ($K_s =
10.3$\,mag) which has been observed with \emph{Spitzer} and eclipses
measured to precisions of 0.011\% and 0.016\% for the two channels
respectively \citep{knutson:2009}.  What HAT-P-24b lacks in a slighter
fainter host star it makes up for with a slightly longer duration than
TrES-4b ($\sim12000$\,s for TrES-4b, see \citet{mandushev:2007}, and
$13150$\,s for HAT-P-24b).  We therefore estimate that the secondary
eclipses will be detectable with SNRs of $\sim7$ and $\sim6$ for
3.6$\mu$m and 4.5$\mu$m respectively.

\subsection{Transmission Spectroscopy}
\label{sec:transmission}

Molecular constituents in the terminator of the atmosphere can absorb
light and cause the transit depth to increase.  The spectral variations
in the transit depth allow for the detection of molecules within an
exoplanet's atmosphere.  This depth change can be estimated from first
principles by calculating the scale height of the atmosphere and
computing the expected change in transit depth using equation (36) from
\citet{winn:2010}.

Using the posterior distribution of parameters from the
\{$b$,$T_{1.5,3.5}$\} fitted-limb-darkening MCMC run, we estimate that
$\Delta \delta \simeq (0.014 \pm 0.001) N_H$\,\% where $N_H$ is the
number of scale heights of atmosphere absorbed by the molecular species
(of order unity), and we have used $\mu_M = 2$\,a.m.u for $H_2$.  Using
the same \emph{Spitzer} uncertainty estimates from before, we would
require $N_H \sim 3$ for even $H_2$ to be detectable.  Therefore,
HAT-P-24b would likely be a challenging target for transmission
spectroscopy.

\subsection{Rossiter-McLaughlin Effect}
\label{sec:rm}

Owing to HAT-P-24's relatively rapid stellar rotation, a large
Rossiter-McLaughlin effect (\citet{rossiter:1924}; 
\citet{mclaughlin:1924}) amplitude is expected. Equally, a large
signal for spectral line tomography could be detectable \citep{cam10}.
Using equation (40) from \citet{winn:2010}, and the posterior
distribution of parameters from the \{$b$,$T_{1.5,3.5}$\}
fitted-limb-darkening MCMC run, we estimate $\Delta V_{RM} = (95 \pm
5)$\,m/s.  Given the RV measurements are essentially jitter-limited at
$7.2$\,m/s, this indicates that we expect a very large signal-to-noise
for the RM effect of HAT-P-24b, reaching SNR$\sim$13.

Another motivation for measuring the planet's RM effect is that
HAT-P-24 has an effective temperature of $T_{\mathrm{eff}} = (6373 \pm
80)$\,K and thus lies above the $\sim6250$\,K threshold for which most
system seem to exhibit significant obliquities \citep{winn:2010b}. 
This therefore indicates that we can not only expect a very large RM
effect but also possibly a highly oblique configuration.

\citet{gaudi:2007} showed that $v \sin i$ and $\lambda$ become
degenerate for low impact parameter transits and so the large RM
amplitude predicted for this system should help in solving for the
system parameters.


\section{Summary}
\label{sec:summary}

We announce the detection of a 0.68\,$M_J$ transiting exoplanet on a
3.36\,day orbit around an F8 star (system parameters are found in the
last column of Table~\ref{tab:global}).  We find that the planet
retains a small eccentricity of $e=0.052_{-0.017}^{+0.022}$ with a
5.8\% false alarm probability, which may suggest either a perturbing
planet in the system or low tidal dissipation within the planet of $Q_P
\gtrsim (6 \pm 3) \times 10^6$.  Most of the eccentricity originates
from the $e\cos\omega$ term and thus we predict that a secondary
eclipse observation, which is shown to be feasible, should
confirm/reject the eccentricity hypothesis conclusively.

We have performed a detailed investigation of the effects upon the
system parameters by using different fitting sets.  In three different
parameter sets, we find a consistent solution indicating the result is
not sensitive to the priors.  The effects of fixing versus fitting limb
darkening coefficients are also investigated, which leads to slightly
increased error bars but a consistent best-fit solution.

Using the Bayesian Information Criterion as a model selection tool, we
find the Keck radial velocities are best described by a model
consisting of non-zero orbital eccentricity and a negative linear drift
of $(-14.6 \pm 10.2)$\,m/s/year with a false alarm probability of 20\%. 
We consider this trend to be currently not statistically significant,
but warranting further investigation.

HAT-P-24 has a relatively rapid stellar rotation of $v \sin i = (10.0
\pm 0.5)$\,km/s, and we therefore predict HAT-P-24b should exhibit one
of the largest known Rossiter-McLaughlin effect amplitudes for an
exoplanet ($\Delta V_{\mathrm{RM}} \simeq 95$\,m/s).  Further, it has
recently been suggested by \citet{winn:2010b} that hot-stars have
companions on preferentially oblique orbits and so HAT-P-24
($T_{\mathrm{eff}} = (6373 \pm 80)$\,K) would be an excellent target to
further investigate this hypothesis.

\acknowledgements 

HATNet operations have been funded by NASA grants NNG04GN74G,
NNX08AF23G and SAO IR\&D grants.  DK was supported by STFC and as an
SAO Predoctoral Fellow.  Work of G.\'A.B.~and J.~Johnson were supported
by the Postdoctoral Fellowship of the NSF Astronomy and Astrophysics
Program (AST-0702843 and AST-0702821, respectively).  GT acknowledges
partial support from NASA grant NNX09AF59G.  We acknowledge partial
support also from the Kepler Mission under NASA Cooperative Agreement
NCC2-1390 (D.W.L., PI).  G.K.~thanks the Hungarian Scientific Research
Foundation (OTKA) for support through grant K-81373.  This research has
made use of Keck telescope time granted through NOAO and NASA\@.  This
paper uses observations obtained with facilities of the Las Cumbres
Observatory Global Telescope.



\end{document}